\documentclass{llncs}

\date{\today}
\author{Sid Mijnders \and Boris de Wilde \and Marijn J. H. Heule\thanks{Supported by the Dutch 
Organization for Scientific Research (NWO) under grant 617.023.611}}
\title{Symbiosis of Search and Heuristics for Random $3$-SAT}
\institute{Delft University of Technology, Delft, The Netherlands}

\usepackage{color}
\usepackage{cite}
\usepackage{amsmath}
\usepackage{amssymb}
\usepackage{graphicx}
\usepackage{algorithm}
\usepackage{algorithmic}
\usepackage{subfigure}
\usepackage{figsize}
\usepackage{verbatim}
\usepackage{multirow}
\usepackage{array}
\usepackage{epsfig}
\usepackage{times}
\usepackage{multicol,color}
\usepackage{rotating}
\usepackage{pstricks}
\usepackage{pst-node}

\newcommand{\mf}{\mathcal{F}}

\newcommand{\xdecision}{x_{\mathrm{decision}}}

\begin{document}

\maketitle

\begin{abstract}
When combined properly, search techniques can reveal the full potential of sophisticated 
branching heuristics. We demonstrate this observation on the well-known class of
random 3-SAT formulae. First, a new branching heuristic is presented, which generalizes
existing work on this class. Much smaller search trees can be constructed by using this heuristic. 
Second, we introduce a variant of discrepancy search, called ALDS. 
Theoretical and practical evidence support that ALDS traverses the search tree in a 
near-optimal order when combined with the new heuristic. Both techniques, search and
heuristic, have been implemented in the look-ahead solver {\sf march}. The SAT 2009 
competition results show that {\sf march} is by far the strongest complete solver on 
random $k$-SAT formulae.
\end{abstract}

\section{Introduction}
Satisfiability (SAT) solvers have a rich history of branching heuristics~\cite{Kul09HBSAT}.
These heuristics are crucial for fast performance. 
They can be split into {\em decision heuristics} and {\em direction heuristics}. The former selects 
for each node in the tree a {\em decision variable} to branch on. These decision heuristics determine 
the size of the search tree. For each free value of the decision variable a child node is created. Direction heuristics 
provide the preferred order in which these child nodes should be visited. In case a search tree contains solutions, 
effective direction heuristics can boost performance.

Two techniques are designed to repair errors made by branching heuristics. First, {\em restart strategies} 
\cite{tail} are used to compensate for ineffective choices made by decision heuristics. A new search tree is created after every 
restart. Second, {\em discrepancy search}~\cite{LDS}, a technique used in Constraint Programming, focuses on 
those parts of the search tree that mostly follow the preference of
the direction heuristics. This technique only changes the order in which the search tree 
is traversed. 

In this paper, we present a new branching heuristic for look-ahead SAT solvers, called the 
{\em recursive weight heuristic}. This heuristic is a generalization of earlier look-ahead evaluation 
heuristics~\cite{Li:97,Li:99,Dubois:01}. Regarding the decision heuristic, we show that on 
random 3-SAT formulae the search tree is significantly smaller compared
to alternative heuristics. Although the heuristic is quite expensive in terms of computational costs, 
performance is clearly improved.

Also, the new direction heuristic results in an observable bias in the distribution of solutions. To capitalize 
on this, we developed a new discrepancy search algorithm, called {advanced limited discrepancy search}
(ALDS), which combines features of improved limited discrepancy search~\cite{ILDS} and depth-bounded
discrepancy search~\cite{DDS}. We provide both theoretical and experimental evidence to show that the
combination of ALDS and the recursive weight heuristic, traverse the tree in a near-optimal order on random 3-SAT
formulae. 

The outline of the paper is as follows: First, we will explain look-ahead heuristics in Section~\ref{sec:lookahead}, 
both the existing work and our new heuristic. In Section~\ref{sec:search}, various search techniques will be discussed.
The focus will be discrepancy search and our variant of this technique. Section~\ref{sec:results} will offer theoretical 
and practical results showing that the combination of ALDS and our heuristic is effective on random 3-SAT 
instances. Finally, we draw some conclusions in Section~\ref{sec:conclusions}.

\section{Look-ahead heuristics}
\label{sec:lookahead}
Most work on branching heuristics in the field of Satisfiability focuses on look-ahead SAT solvers~\cite{Kul09HBSAT}.
In contrast to many other solvers, look-ahead solvers keep track of various statistical measurements that make it possible
to use quite complex heuristics. In this section, we first will provide an overview of look-ahead SAT solvers. Afterwards, the branching heuristics in these solvers are discussed. We conclude this section by introducing an improved heuristic.

\subsection{Look-ahead SAT solvers}

The look-ahead architecture for SAT solvers is based on the {\em DPLL} framework~\cite{Davis:1962}: It is a
complete solving method which selects in each step a decision variable $\xdecision$
and recursively calls DPLL for the reduced formula where $\xdecision$ is assigned to false
(denoted by $\mf[\xdecision = 0]$) and another where $\xdecision$ is assigned to true
(denoted by $\mf[\xdecision = 1]$).

A formula $\mf$ is reduced by {\em unit propagation}: Given a formula $\mf$, an unassigned
variable $x$ and a Boolean value {\bf B}, first $x$ is assigned to {\bf B}. If this assignment
$\varphi$ results in a {\em unit clause} (clause of size 1) then $\varphi$ is expanded by assigning the
remaining literal of that clause to true. This is repeated until no unit clauses are
left in $\varphi$ applied to $\mf$. We denote by $\varphi * \mf$ the reduced formula after unit propagation 
of applying $\varphi$ on $\mf$, with all satisfied clauses removed. 
So, more specific than above, $\mf[x\,=\,${\bf B}$] := \varphi * \mf$.

\begin{algorithm}[!ht]
    \caption{ {\sc UnitPropagation}( formula $\mathcal{F}$, variable $x$, {\bf B}$~\in\{0,1\}$ )}
	\label{alg:up}
	\begin{algorithmic}[1]
	\STATE $\varphi$ := $\{x \leftarrow ${\bf B}$\}$
	\WHILE{empty clause $\notin$ $\varphi$ applied on $\mf$ {\bf and} unit clause $(y)$ $\in$ $\varphi$ applied on $\mf$}
	\STATE $\varphi$ := $\varphi \cup \{y \leftarrow 1\}$
	\ENDWHILE
	\STATE {\bf return} $\varphi * \mf$
    \end{algorithmic}
\end{algorithm}

The recursion has two kinds of leaf nodes: Either all clauses have been satisfied (denoted
by $\varphi * \mathcal{F}$ = $\emptyset$), meaning that a satisfying assignment has been found, or
$\varphi * \mf$ contains an {\em empty clause} (a clause of which all literals have been falsified),
meaning a dead end. In the latter case the algorithm backtracks.

\begin{figure}[ht]

\centering
\begin{pspicture}(7,8)

\psset{xunit=17pt, yunit=20pt}

\cnodeput(6.5,10){top}{$x_a$}

\cnodeput(3.5,8.5){dab}{$x_b$}
\cnodeput(9.5,8.5){dcd}{$x_c$}

\ncline{dab}{top}
\ncline{dcd}{top}

\rput(4.5, 9.5){{\tt 0}}
\rput(8.5, 9.5){{\tt 1}}

\cnodeput[fillstyle=solid, fillcolor=black](2,7){da}{~}
\cnodeput[fillstyle=solid, fillcolor=black](5,7){db}{~}
\cnodeput(8,7){dc}{?}

\rput(2.3, 7.8){{\tt 1}}
\rput(4.7, 7.8){{\tt 0}}
\rput(8.3, 7.8){{\tt 0}}

\ncline{da}{dab}
\ncline{db}{dab}
\ncline{dc}{dcd}

\psframe[fillstyle=none](-2,6)(16,11)
\rput[l](-1.5,10.5){{\sc DPLL}}

\cnodeput(1,1.5){a}{}
\cnodeput(3,1.5){b}{}
\cnodeput(5,1.5){c}{}
\cnodeput(7,1.5){d}{}
\cnodeput[fillstyle=solid, fillcolor=black](9,1.5){e}{}
\cnodeput(11,1.5){f}{}
\cnodeput(13,1.5){g}{}
\cnodeput(15,1.5){h}{}

\cnodeput(2,3.5){ab}{$x_1$}
\cnodeput(6,3.5){cd}{$x_2$}
\cnodeput(10,3.5){ef}{$x_3$}
\cnodeput(14,3.5){gh}{$x_4$}

\cnodeput(8,4.5){la}{$\mathcal{F}_{\mathrm{LA}}$}

\ncline{a}{ab} \rput(1.1, 2.5){{\tt 0}}   \rput(1,0.5){3}
\ncline{b}{ab} \rput(2.9, 2.5){{\tt 1}}   \rput(3,0.5){1}
\ncline{c}{cd} \rput(5.1, 2.5){{\tt 0}}   \rput(5,0.5){2}
\ncline{d}{cd} \rput(6.9, 2.5){{\tt 1}}   \rput(7,0.5){2}
\ncline{e}{ef} \rput(9.1, 2.5){{\tt 0}}   \rput(9,0.5){}
\ncline{f}{ef} \rput(10.9, 2.5){{\tt 1}}  \rput(11,0.5){1}
\ncline{g}{gh} \rput(13.1, 2.5){{\tt 0}}  \rput(13,0.5){2}
\ncline{h}{gh} \rput(14.9, 2.5){{\tt 1}}  \rput(15,0.5){1}

\ncline{ab}{la}
\ncline{cd}{la}
\ncline{ef}{la}
\ncline{gh}{la}

\ncline{dc}{la}

\psframe[fillstyle=none](-4,0)(16,5.5)
\rput[l](-3.5,5){{\sc LookAhead}}

\rput[l](-3.5,0.5){$|\mathcal{C}_{\mathrm{new}}|$}

\end{pspicture}
\caption{A graphical representation of the look-ahead architecture. Above, the DPLL
super-structure (a binary tree) is shown. In each node of the DPLL-tree, the {\sc LookAhead}
procedure is called to select the decision variable and to compute implied variables by
additional reasoning. Black nodes refer to leaf nodes.}
\label{fig:examla}
\end{figure}
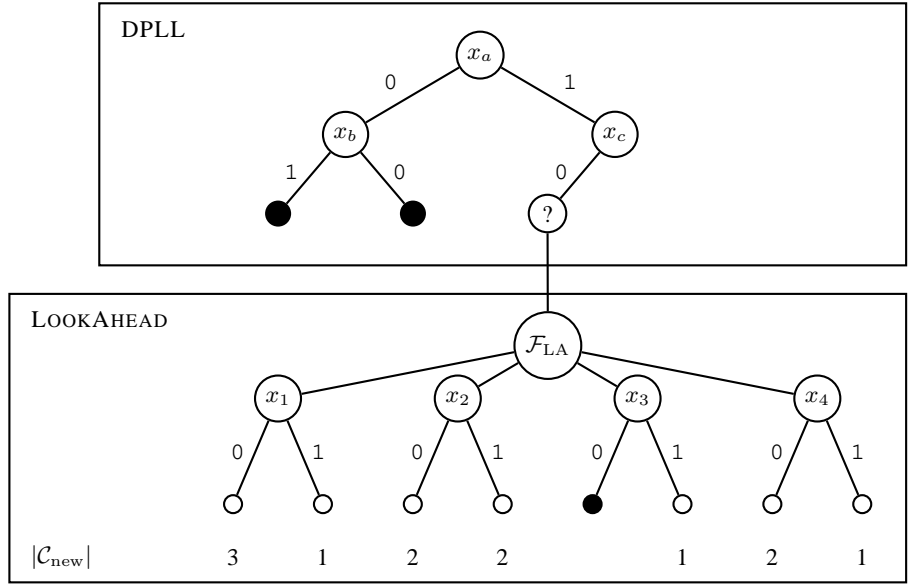

The core of the look-ahead architecture is the {\sc LookAhead} procedure, which incorporates  
the branching heuristics (selecting a decision variable and selecting the first branch) 
and several reasoning techniques to reduce the size of the formula.
Because the latter are beyond the scope of this paper, we refer the reader to~\cite{HvM09HBSAT}
for details. Algorithm~\ref{alg:dpll} shows the top level structure. 
Notice that the {\sc LookAhead} procedure returns a reduced formula $\mathcal{F}$, variable $\xdecision$, 
and value {\bf B}. Fig.~\ref{fig:examla} provides a graphical overview of the architecture.

\begin{algorithm}[!ht]
    \caption{ {\sc DPLL}( formula $\mathcal{F}$ )}
	\label{alg:dpll}
    \begin{algorithmic}[1]
    \IF{$\mathcal{F}$ = $\emptyset$}
    \STATE {\bf return} {\tt satisfiable}
    \ENDIF
    \STATE $<\mathcal{F}; ~x_{\rm decision};~${\bf B}$\,>$ := {\sc LookAhead}( $\mathcal{F}$ )
    \IF{empty clause $\in \mathcal{F}$}
    \STATE {\bf return} {\tt unsatisfiable}
	\ENDIF
    \IF{ DPLL( $\mathcal{F}[x_{\rm decision}$ = {\bf B}] ) = {\tt satisfiable}}
    \STATE {\bf return} {\tt satisfiable}
    \ENDIF
    \STATE {\bf return} DPLL( $\mathcal{F}[x_{\rm decision}$ = $\lnot${\bf B}] )
    \end{algorithmic}
\end{algorithm}

The {\sc LookAhead} procedure, as the name suggests, performs {\em look-aheads}.
A look-ahead on $x$ starts by assigning $x$ to true followed by unit propagation.
The importance of $x$ is measured {\em and} possible reductions of the formula are
detected. After this analysis, it backtracks, ending the look-ahead.
The rationale of a look-ahead operation is that evaluating the effect of actually
assigning variables to truth values and performing unit propagation is more
adequate than taking a cheap guess using statistical data on $\mf$.

\subsection{Look-ahead evaluation}

Branching heuristics in look-ahead SAT solvers are based on evaluating the
reduction of the formula during a look-ahead. This reduction is expressed using
the {\em difference} or {\em distance heuristic} (in short {\sc Diff}). The larger the 
reduction, the higher the heuristic value. A {\sc Diff} could be based
on many statistics, such as the reduction of the number of variables. Yet, all 
look-ahead SAT solvers use a  {\sc Diff} based only on the set of newly created
(i.e. reduced, but not satisfied) clauses, denoted by $\mathcal{C}_{\mathrm{new}}$~\cite{HvM09HBSAT}.

The decision variable $\xdecision$ is selected by combining for each variable $x_i$
the values {\sc Diff}($\mf$, $\mf$[$x_i$ = 0]) and {\sc Diff}($\mf$, $\mf$[$x_i$ = 1]). 
The objective of the decision heuristic is to construct a small and balanced 
search tree. The product of these numbers is generally considered
to be an effective heuristic for this purpose~\cite{Kul09HBSAT}. The sum
of these numbers can be used for tie-breaking.

Once $\xdecision$ is selected, the direction heuristics decide whether
to branch first on $\mf$[$\xdecision$ = 0] or $\mf$[$\xdecision$ = 1].
Most solvers prefer the branch which is the most satisfiable~\cite{Kul09HBSAT,HvM09HBSAT}.
A heuristic used to determine the most satisfiable branch selects Boolean value {\bf B} 
for which 
{\sc Diff}($\mf$, $\mf$[$x_i$ = {\bf B}]) is the smallest.

Consider the following example formula:

\[
\mathcal{F}_{\mathrm{LA}} = (\lnot x_1 \lor x_3) \land (x_1 \lor x_2 \lor  x_3)
    \land (x_1 \lor \lnot x_2 \lor x_4) \land (x_1 \lor \lnot x_2 \lor \lnot x_4) \land
        ( x_2 \lor \lnot x_3 \lor x_4 )
\]

Since all clauses in $\mathcal{F}_{\mathrm{LA}}$
have size three or smaller, only new binary clauses can be created. For instance, during the
look-ahead on $\lnot x_1$, three new binary clauses are created (all clauses in
which literal $x_1$ occurs). The look-ahead on $x_1$ will force $x_3$ to
be assigned to true by unit propagation. This will reduce the last clause to a binary
clause, while all other clauses become satisfied. Similarly, we can compute the number
of new binary clauses for all look-aheads -- see Fig.~\ref{fig:examla}.

Finally, the selection of the decision variable is based on the reduction measurements of
both the look-ahead on $\lnot x_i$ and $x_i$. Generally, the product is used
to combine the numbers. In this example, $x_2$ would be selected as decision variable, because
the product of the reduction measured while performing look-ahead on $\lnot x_2$ and $x_2$ is
the highest (i.e. 4). 

The {\sc Diff} heuristic based on the number of newly created clauses was introduced by Li and 
Anbulagan~\cite{Li:97}. Several extensions have been proposed dealing with how to weigh clauses
in $\mathcal{C}_{\mathrm{ new}}$. If the original formula is $3$-SAT, $\mathcal{C}_{\mathrm{ new}}$
only consists of binary clauses. For this special case, Li~\cite{Li:99} uses in {\sf satz} weights based
on the occurrences of variables. Let $\#(x)$ denote the number of occurrences of literal $x$. 
Each clause $(y \lor z) \in \mathcal{C}_{\mathrm{ new}}$ gets a weight of 
$\#(\lnot y) + \#(\lnot z)$. A slight variation but much more effective weight is used by Dubois
and Dequen~\cite{Dubois:01} in their solver {\sf kcnfs}. Their backbone search heuristic weighs
a binary clause $(y \lor z)$ by $\#(\lnot y) \times \#(\lnot z)$. In case of $k$-SAT instances, 
Kullmann~\cite{Kullmann:2002} uses in the {\sf OKsolver} weights based on the length of clauses in
$\mathcal{C}_{\mathrm{ new}}$. A clause of size $k$ roughly get a weight of 
$\gamma_k = 5^{2-k}$.

\subsection{Recursive weight heuristic}
\label{sec:rwh}

We developed a model to generalize existing work on look-ahead branching heuristics.
We refer to this model as the {\em recursive weight heuristic}. Let $V\! AR(\mathcal{F})$ refer to the set of variables 
in $\mathcal{F}$ and $n = |V\! AR(\mathcal{F})|$. 
The heuristic values $h_i(x)$ express for each iteration $i$ how much literal $x$ is forced to true by the clauses containing $x$.
First, for all literals $x$, $h_0(x)$ are initialized on 1:

\begin{equation}
h_0(x) = h_0(\lnot x) = 1
\end{equation}

At each step, the heuristics values $h_i(x)$ are scaled using the average value $\mu_i$:

\begin{equation}
\mu_i = \frac{1}{2n} \sum_{x \in V\! AR(\mathcal{F})} \big( h_i(x) + h_i(\lnot x) \big)
\end{equation}

Finally, in each next iteration, the heuristic values $h_{i+1}(x)$ are computed in which literals 
$y$ get weight ${h_{i}(\lnot y)}/{\mu_i}$. Weight $\gamma$ expresses the relative importance of binary 
clauses. This weight could also be seen as the heuristic value of a falsified literal.

\begin{equation}
h_{i+1}(x) = \sum_{(x \lor y \lor z) \in \mathcal{F}} \Big(\frac{h_i(\lnot y)}{\mu_i} \cdot \frac{h_i(\lnot z)}{\mu_i}\Big) + 
			\gamma \!\!\!\! \sum_{(x \lor y) \in \mathcal{F}} \frac{h_i(\lnot y)}{\mu_i}
\end{equation}

Earlier work on look-ahead heuristics can be formulated using the model above. For
binary clause $(y \lor z) \in \mathcal{C}_{\mathrm{new}}$ they compute a weight $w^*_i(y\lor z) = h_i (\lnot y)~*~h_i (\lnot z)$:

\begin{itemize}
\item Li \& Anbulagan 1997~\cite{Li:97}: $w(y \lor z) = 1 = h_0 (\lnot y) \times  h_0 (\lnot z)$, 
								in short $w_0^{\times}$.\\ \vspace{-5pt}
\item Li 1999~\cite{Li:99}: $w(y \lor z) = \#(\lnot y) + \#(\lnot z) = h_1 (\lnot y) +  h_1 (\lnot z)$,
								 in short $w_1^{+}$.\\ \vspace{-5pt}
\item Dubois 2001~\cite{Dubois:01}: $w(y \lor z) = \#(\lnot y) \times \#(\lnot z) = h_1 (\lnot y) \times  h_1 (\lnot z)$,
								 in short $w_1^{\times}$.
\end{itemize}

Although~\cite{Li:99,Dubois:01} used $\gamma=5$, we observed stronger performance using $\gamma=3.3$.
Also, the size of the tree can be reduced significantly by using weights $w_i^{\times}$ with $i > 1$.

\begin{figure}[ht]
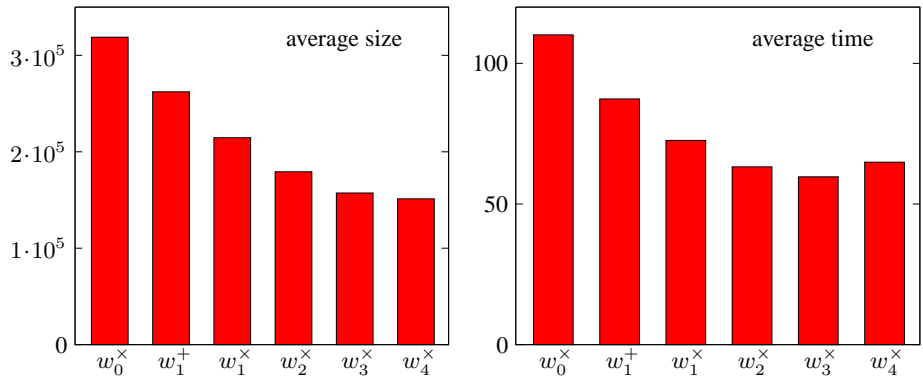

\begin{minipage}{0.45\textwidth}
\include{ps2_size} 
\end{minipage}
\hfill
\begin{minipage}{0.45\textwidth}
\include{ps2_time}
\end{minipage}
\caption{The influence of different heuristics for the clause weight used for look-ahead evaluation on the 
average size of the search tree and the computational costs. The results are average sizes and times for 500 
random 3-SAT formulae with 450 variables and 1915 clauses.}
\label{fig:results}
\end{figure}

We implemented $w_0^{\times}$, $w_1^{+}$, $w_1^{\times}$, $w_2^{\times}$, $w_3^{\times}$ and $w_4^{\times}$ 
in the look-ahead solver {\sf march\_ks} \cite{side} with $\gamma = 3.3$. We experimented 
on 500 random 3-SAT formulae with 450 variables and 1915 clauses (phase transition density).
To provide stable numbers, all instances were unsatisfiable. Fig.~\ref{fig:results} shows the
results. Clearly, the average size of the tree is smaller using $w_3^{\times}$ compared to the
alternative heuristics. Although $w_3^{\times}$ is much more expensive to compute, the average
time to solve these instances has also decreased. Regarding the computational costs, we observed
that $w_3^{\times}$ resulted in best performance of all $w_i^{\times}$ on random 3-SAT formulae.
In case $i>3$, the reduction of the size of the tree is not large enough to compensate for the 
additional cost to compute the weights. 

Our implementation with $w_3^{\times}$ participated as {\sf march\_hi} at the SAT competition of 2009\footnote{see {\tt \url{http://www.satcompetition.org}} for details}. It won the
random unsatisfiable category. Apart from a few minor optimizations and fixes, the only difference compared 
to {\sf march\_ks} (the winner in 2007) is the recursive weight heuristic. Both versions competed during SAT 2009 and {\sf march\_hi} solved over 10\% more unsatisfiable instances. For many instances in this category it was the only program to solve 
them.

\section{Heuristic search}
\label{sec:search}
Let us take a step back from SAT to consider how to capitalize on direction heuristics in general.
We will first discuss the terminology of direction heuristics, and the ideas behind the most important 
heuristic search strategies. Afterwards, we will introduce an alternative search strategy, which is very 
powerful in combination with the recursive weight heuristic. 

\subsection{Direction heuristics}
\label{section-heuristic}
\def \goal {\mathrm{goal\mbox{-}node}}
\def \leftchild {\mathrm{left\mbox{-}child}}
\def \rightchild {\mathrm{right\mbox{-}child}}
\def \depth {\mathrm{depth}}
\def \cond {\;|\;}
\def \prob [#1] {P(\:#1\:)}
\def \pgoal {P_{\mathrm{goal}}}
\def \pheur {P_{\mathrm{heur}}}

There are various ways to explore search trees.
Searching the entire tree for a specific {\em goal node} is costly. 
Therefore, {\em search strategies} have been developed to guide the search towards a goal node.
To show that a problem has no solutions, the search has to be \emph{complete} by visiting
all leaf nodes.
Complete search strategies will either detect a goal node or prove that none exists.

If a search tree contains goal nodes, {\em direction heuristics} predict which branches 
have a higher probability of leading to a goal node than others.
The branch with the highest preference will be called the {\em left branch}.
Any other branch is called a \emph{discrepancy}.
In the case of a binary search tree a discrepancy can also be referred to as the \emph{right branch}.

In theory, direction heuristics are very powerful. Perfect direction heuristics would lead to the goal node immediately.
If such a perfect direction heuristic would exist, given that it is computable in polynomial time, 
it would prove that $\mathcal{P}=\mathcal{NP}$.
At each node direction heuristics have a probability of picking the correct branch as left branch.
This probability is called the \emph{heuristic probability}.
Another probability that we define is the \emph{goal node probability}.
For each node it expresses the probability that the subtree with this node as root contains a goal node.
The heuristic probability $\pheur(v)$ and the goal node probability $\pgoal(v)$ are related as follows:
\begin{eqnarray*}
\pgoal(v) & = & \prob[\goal(v)] \\
\pheur(v) & = & \prob[\goal(\leftchild(v)) \cond \goal(v)] \\
\pgoal(\leftchild(v)) & = & \prob[\goal(\leftchild(v))] \\
 & = & \pgoal(v)\cdot \pheur(v)
\end{eqnarray*}
In the case of the binary search tree we can also define:
\begin{eqnarray*}
\pgoal(\rightchild(v)) & = & \prob[\goal(\rightchild(v))] \\
 & = & \pgoal(v)\cdot (1-\pheur(v))
\end{eqnarray*}

For a given dataset $D$ and search strategy $S$, $\pgoal(v)$ denotes the fraction of formulae in $D$ 
that contain solutions in the subtree rooted at $v$, while applying \mbox{algorithm $S$}.

It is observed \cite{side}, that heuristics tend to make more mistakes in the top of the search tree.
As we get closer to the leaf nodes in the tree, the underlying problem has been simplified.
Heuristics perform better on  simplified problems.
Therefore, it is expected that $\pheur(v)$ 
increases while descending in the search tree.
Consequently, direction heuristics are most likely to make mistakes near the root of the search tree.

\begin{figure}[ht]
\centering
\includegraphics[width=0.75\textwidth]{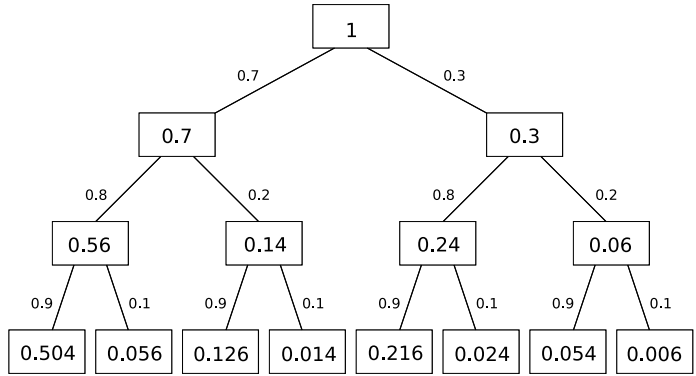}
\caption{Binary search tree containing one goal node.
Nodes are labelled with the goal node probability $\pgoal(v)$.
Edges show the heuristic probabilities $\pheur(v)$ (left) and $1-\pheur(v)$ (right).}
\label{prob-tree}
\end{figure}

To illustrate  goal node probabilities throughout a tree, we will consider 
a direction heuristic with increasing $\pheur(v)$ in a binary tree with a single solution, see Fig.~\ref{prob-tree}.
Notice that, if a problem has solutions, $\pgoal(\mathrm{root})=1$.

Looking at the goal node probabilities in this example, it shows that, although for each node individually the left child has a higher $\pgoal(v)$ than the right child, when comparing all nodes at a certain depth no clear pattern can be observed between the $\pgoal(v)$ values. 
While searching for the goal node, one wants to take the $\pgoal(v)$ of the leaf nodes into account.
In this example, when a search strategy visits the nodes with a high goal node probability first, it will on average visit 
less leaf nodes before finding a goal node.
We will now discuss several complete search strategies.

\subsection{Depth first search}
\begin{figure}[ht]
\centering
$\begin{tabular*}{\textwidth}{@{}@{\extracolsep{\fill}} c c c c @{}}
~\\
\multicolumn{1}{l}{\mbox{\scriptsize 1}}  &
\multicolumn{1}{l}{\mbox{\scriptsize 2}}  &
\multicolumn{1}{l}{\mbox{\scriptsize 3}}  &
\multicolumn{1}{l}{\mbox{\scriptsize 4}} \\ [-0.5cm]
\includegraphics[width=0.23\textwidth]{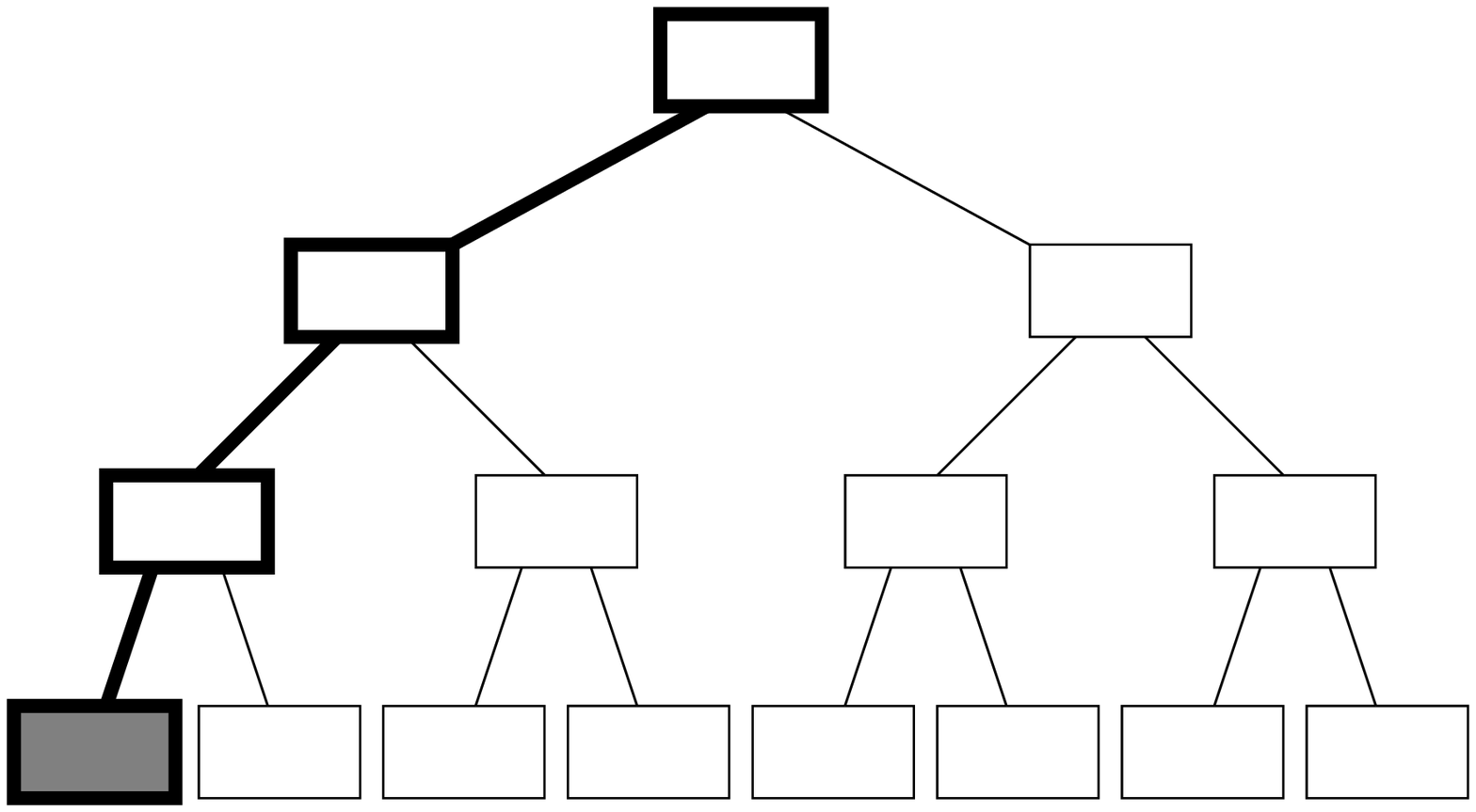} &
\includegraphics[width=0.23\textwidth]{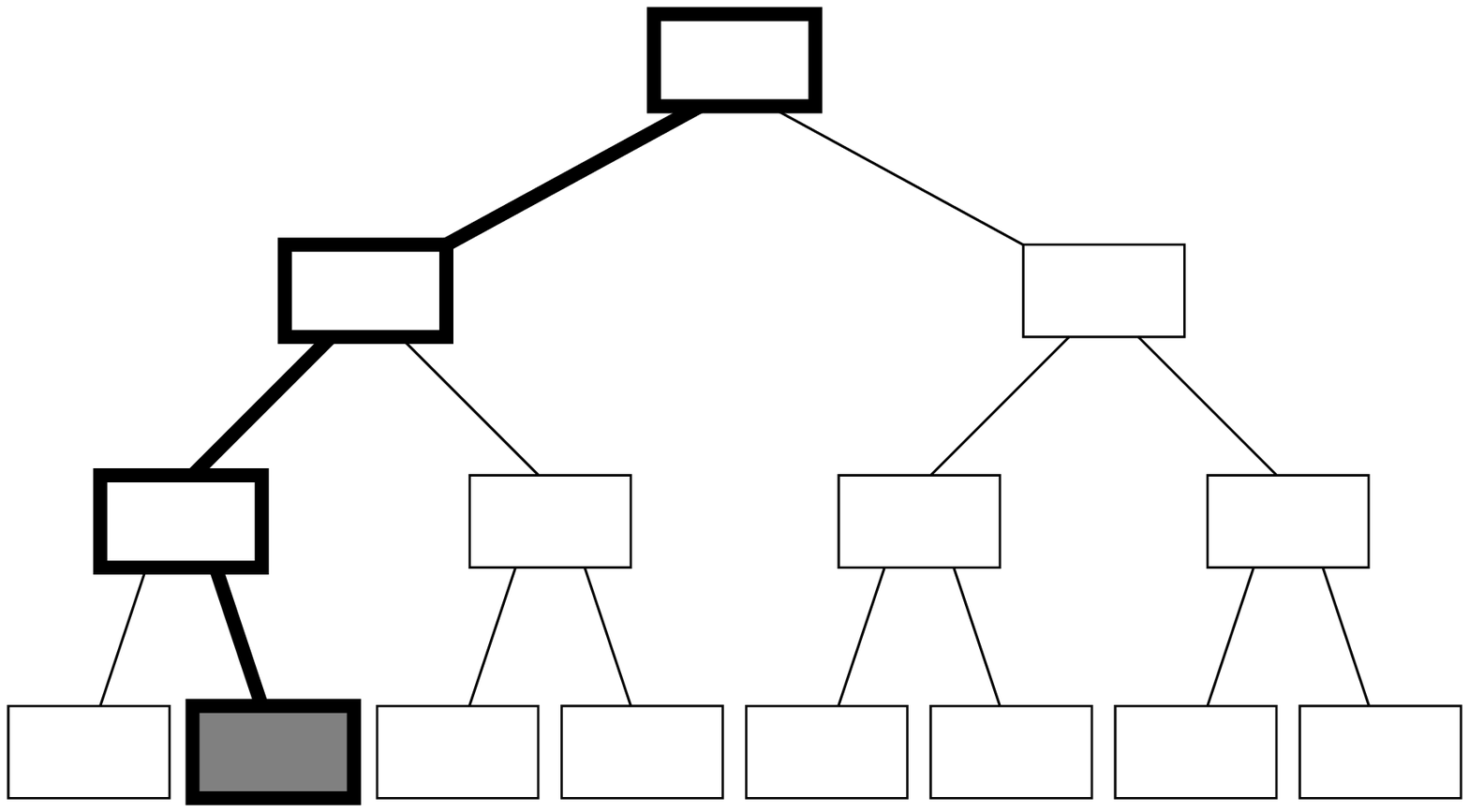} &
\includegraphics[width=0.23\textwidth]{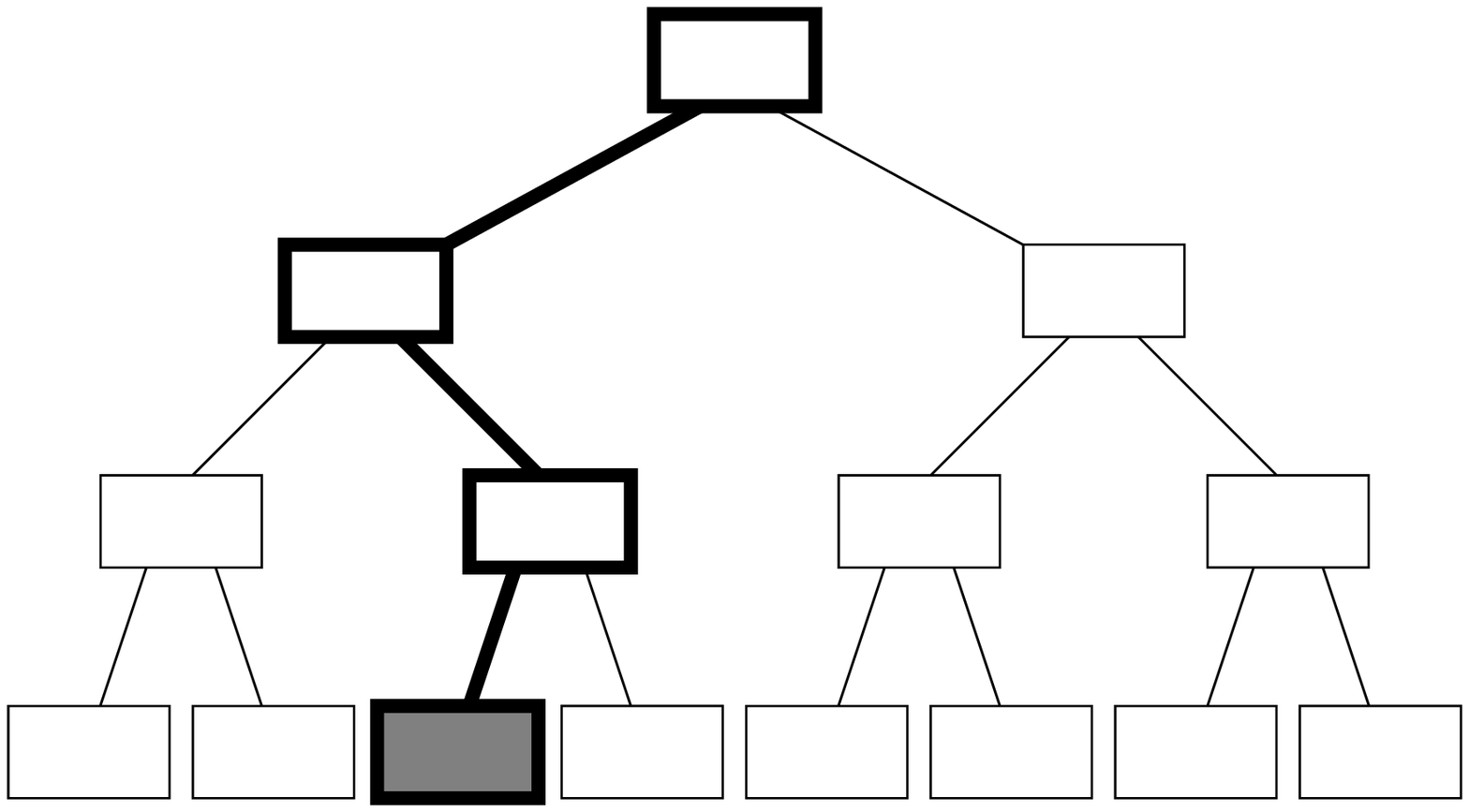} &
\includegraphics[width=0.23\textwidth]{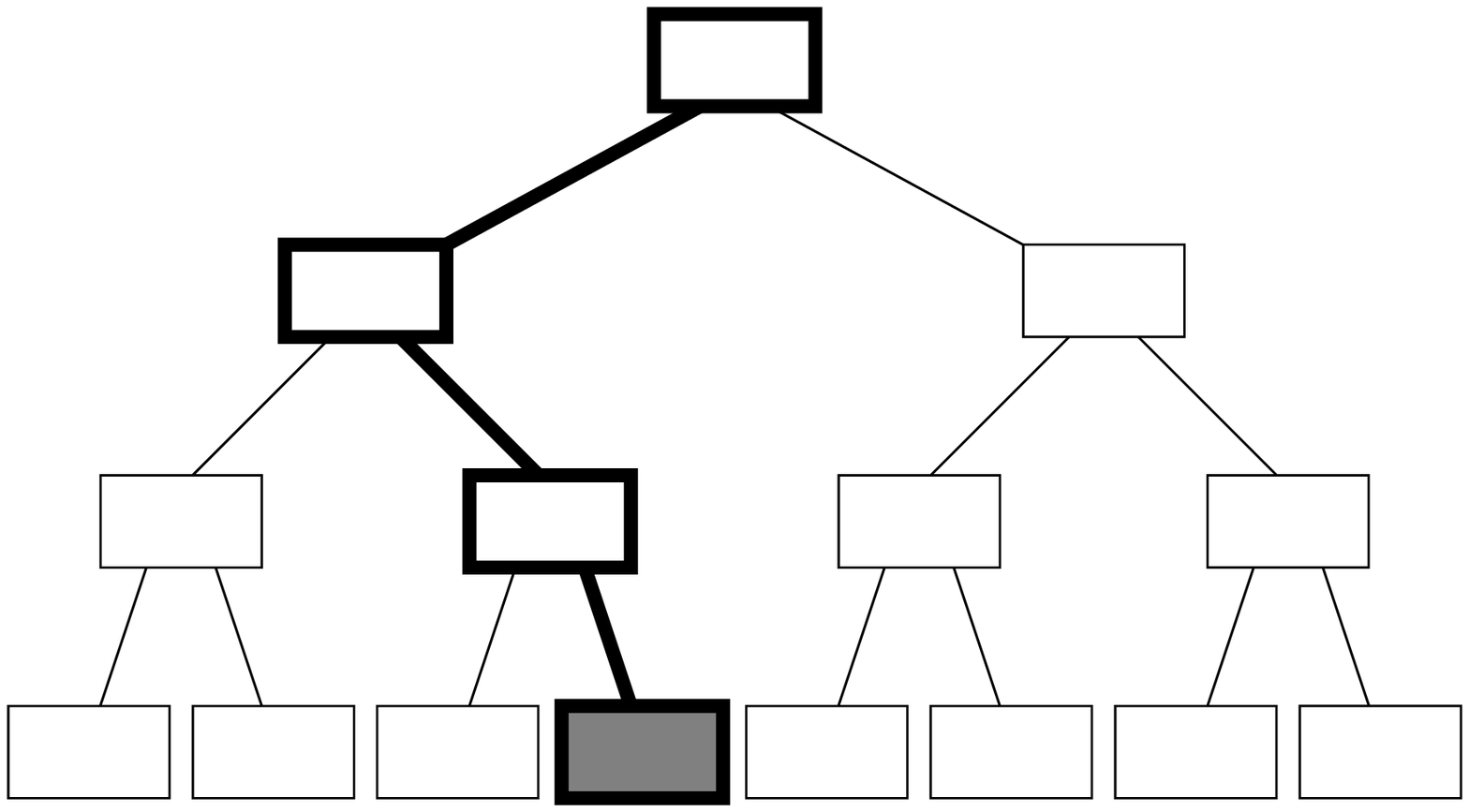} \\ [0.3cm]

\multicolumn{1}{l}{\mbox{\scriptsize 5}}  &
\multicolumn{1}{l}{\mbox{\scriptsize 6}}  &
\multicolumn{1}{l}{\mbox{\scriptsize 7}}  &
\multicolumn{1}{l}{\mbox{\scriptsize 8}}  \\ [-0.5cm]
\includegraphics[width=0.23\textwidth]{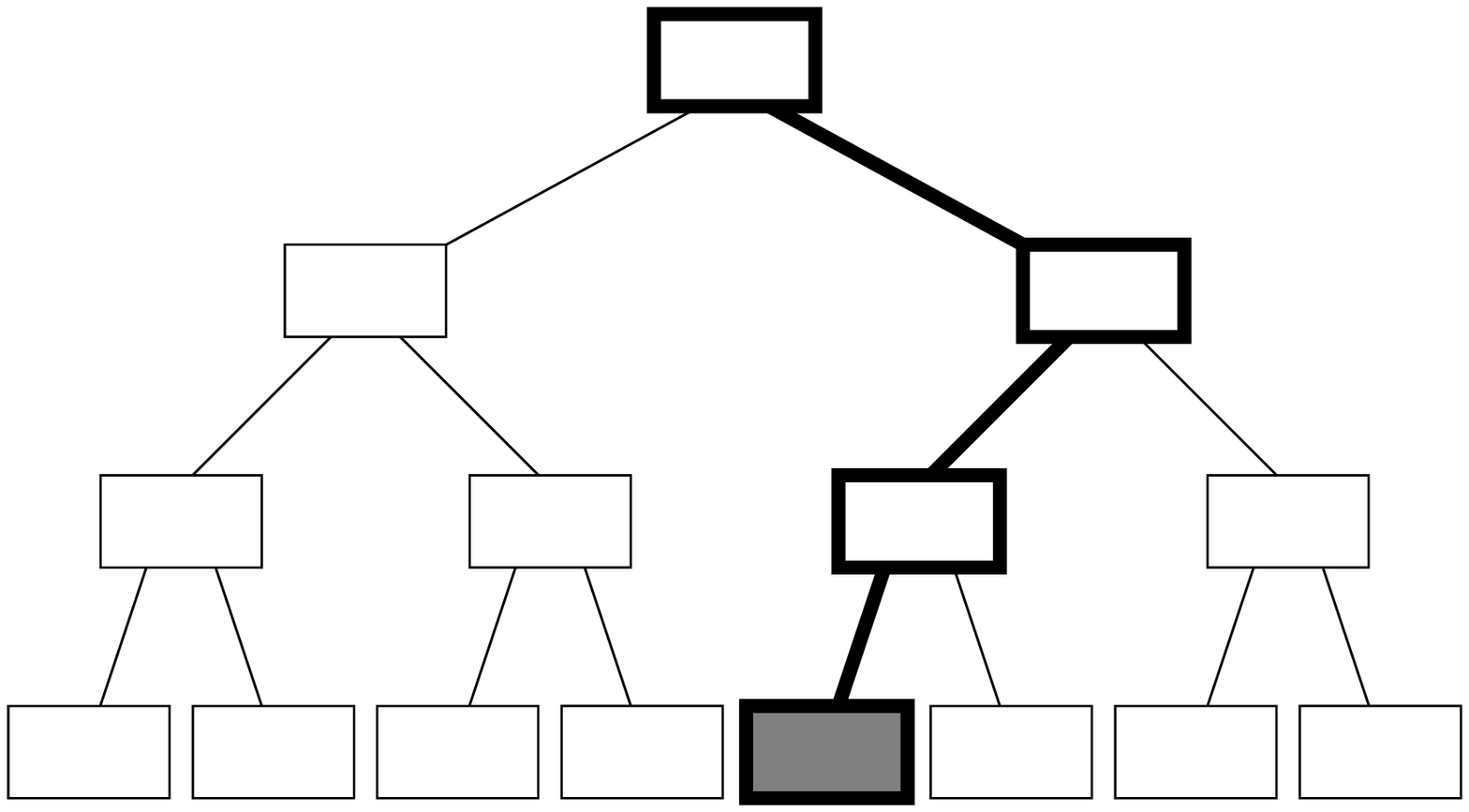} &
\includegraphics[width=0.23\textwidth]{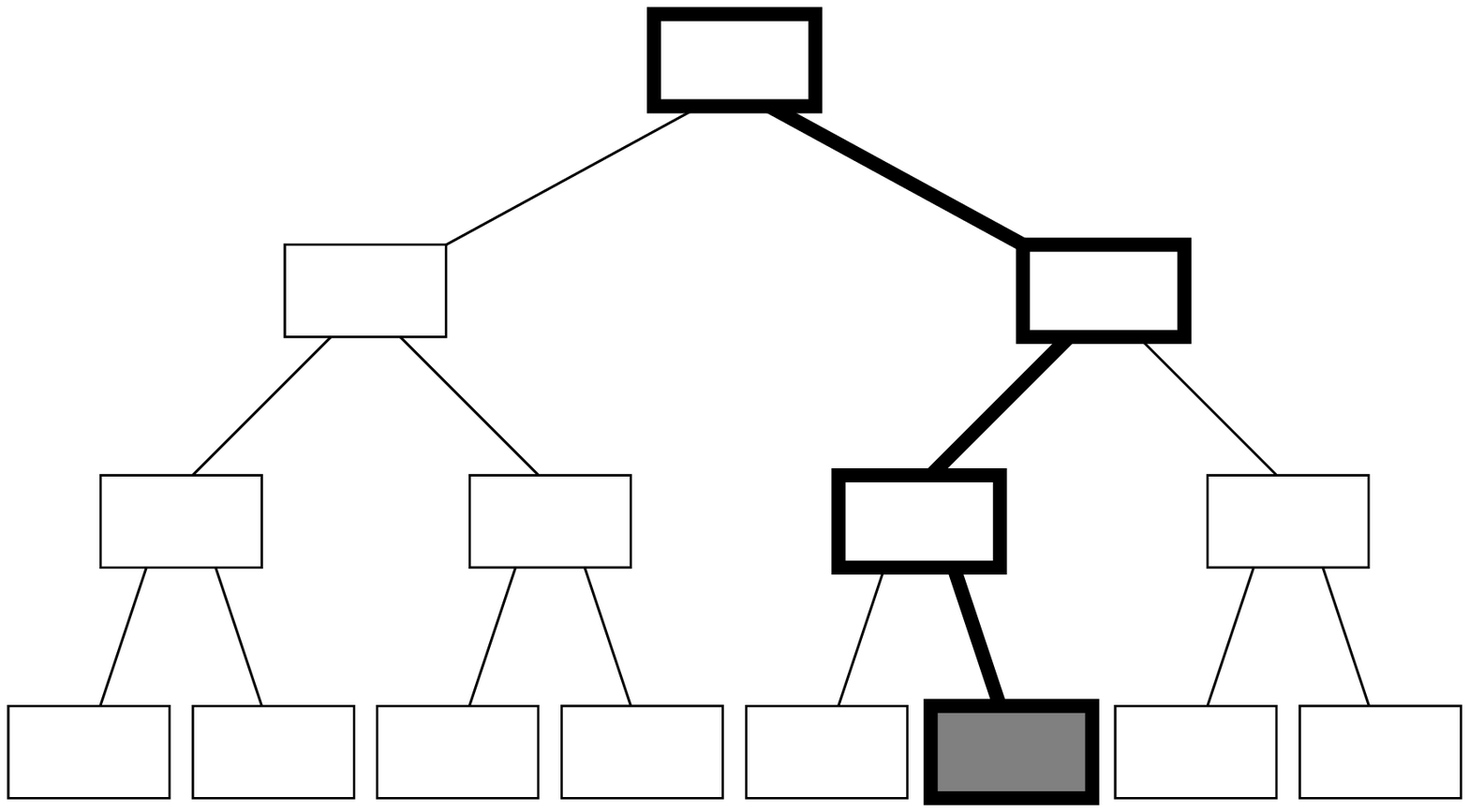} &
\includegraphics[width=0.23\textwidth]{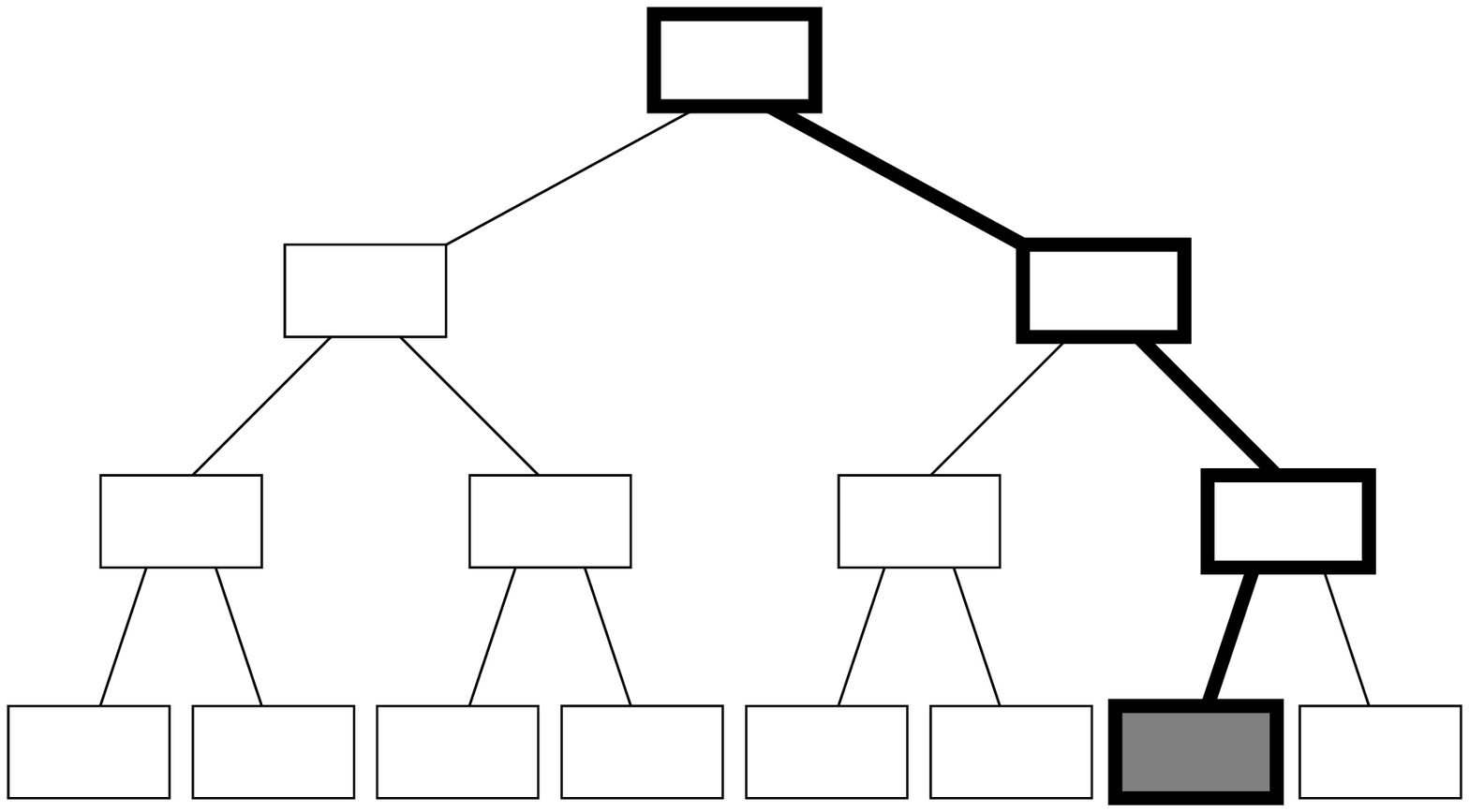} &
\includegraphics[width=0.23\textwidth]{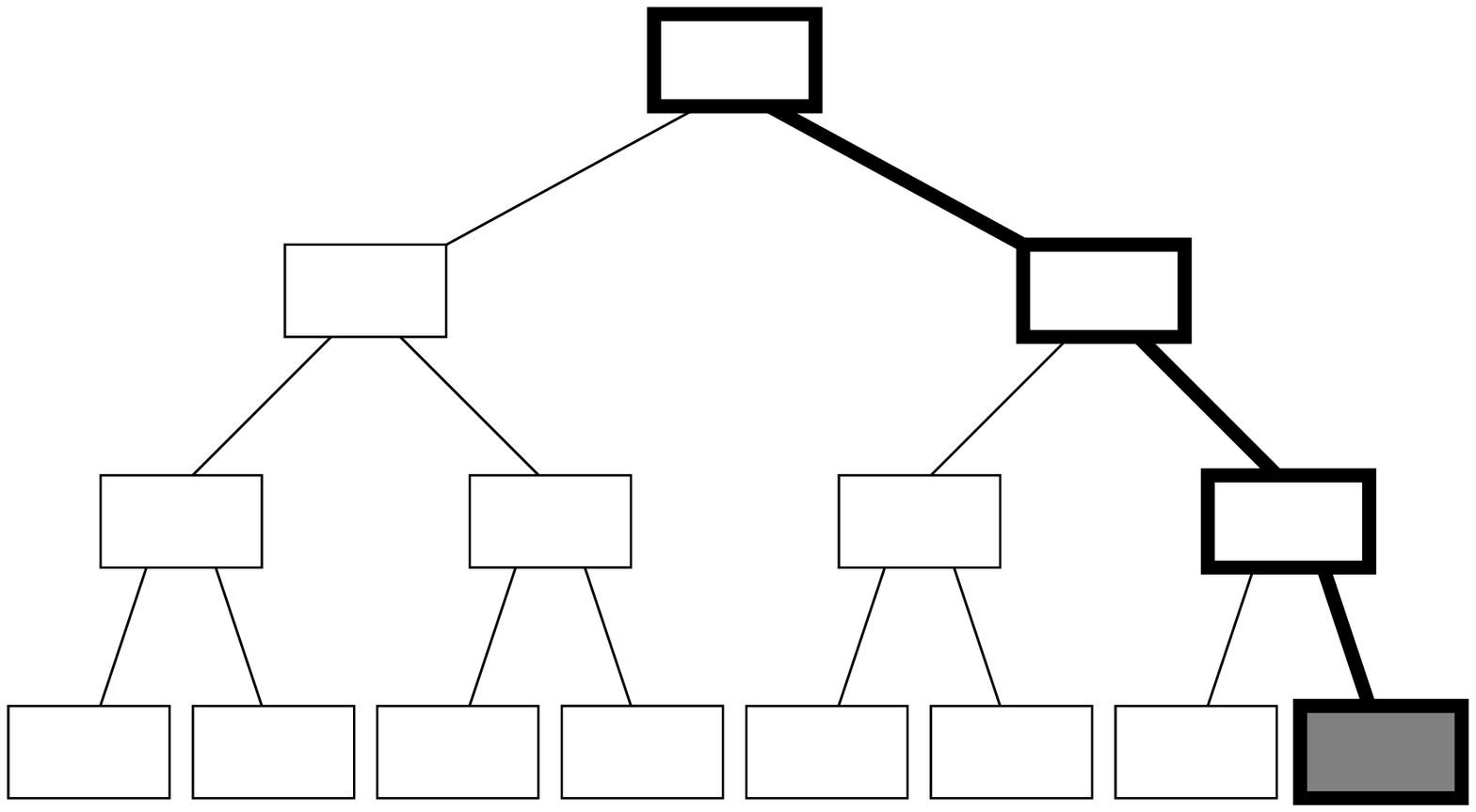} 
\end{tabular*}$
\caption{The iterations of DFS on a search tree of depth 3.}
\label{dfs-iterations}
\end{figure}

One of the best know search strategies is \emph{depth first search} (DFS). 
DFS branches left until it reaches a leaf node, after which it backtracks chronologically.
The order in which DFS visits leaf nodes, from left to right, is shown in Fig.~\ref{dfs-iterations}.
DFS traverses the minimum number of edges needed to explore the entire tree.
When $\pheur(v)=0.5$ for all nodes $v$ in a binary tree, meaning that the direction heuristic might as well randomly select branches, $\pgoal(v)$ is equal for every node $v$ at the same depth, then DFS is the cheapest strategy to use.
When direction heuristics are stronger than random selection, as in Fig.~\ref{prob-tree}, 
alternative strategies traverse the tree more efficiently.
%

\subsection{Discrepancy search}
Assume that direction heuristics perform significantly better than random branch selection.
When some leaf node is not a goal node, the intuition is that the heuristic only took a small number of wrong branches on the path to this leaf node.
Intuitively, tree search performs best when paths with a small number of discrepancies are explored first.

\subsubsection{Limited Discrepancy Search (LDS)}  
\begin{figure}[ht]
\centering
$\begin{tabular*}{\textwidth}{@{}@{\extracolsep{\fill}} c c c c @{}}
~\\
\multicolumn{1}{l}{\mbox{\scriptsize 1}}  &
\multicolumn{1}{l}{\mbox{\scriptsize 2}}  &
\multicolumn{1}{l}{\mbox{\scriptsize 3}}  &
\multicolumn{1}{l}{\mbox{\scriptsize 4}} \\ [-0.5cm]
\includegraphics[width=0.23\textwidth]{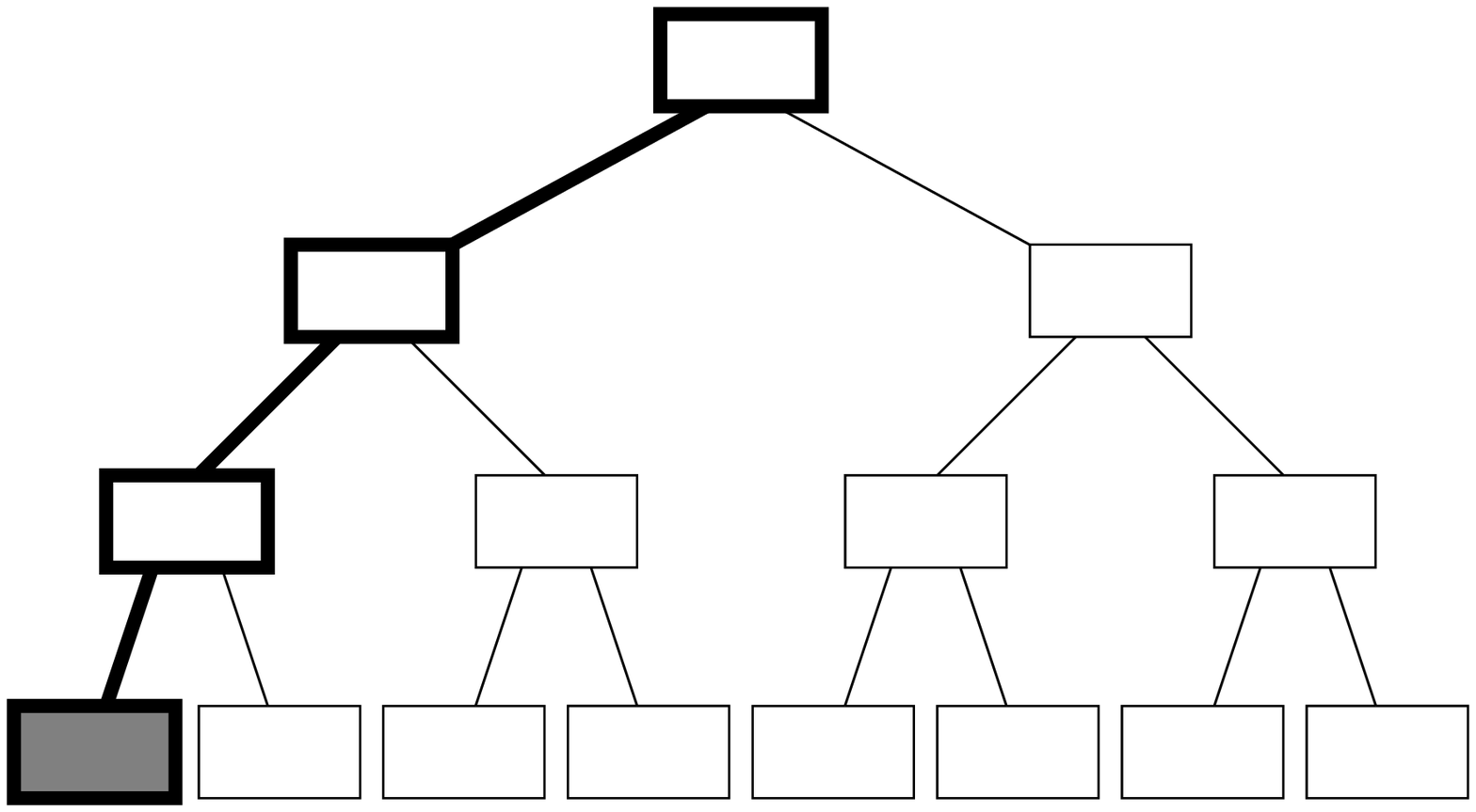} &
\includegraphics[width=0.23\textwidth]{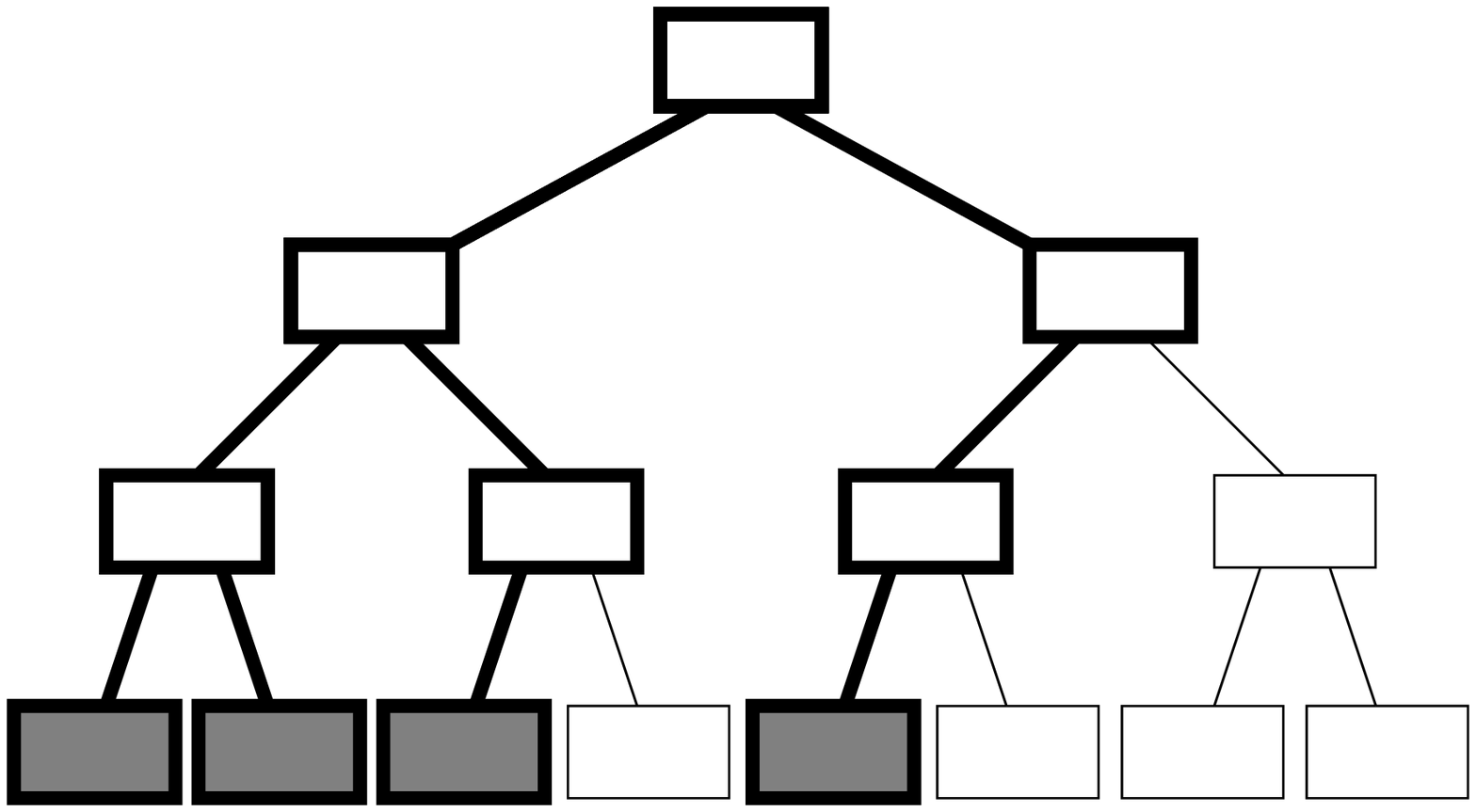} &
\includegraphics[width=0.23\textwidth]{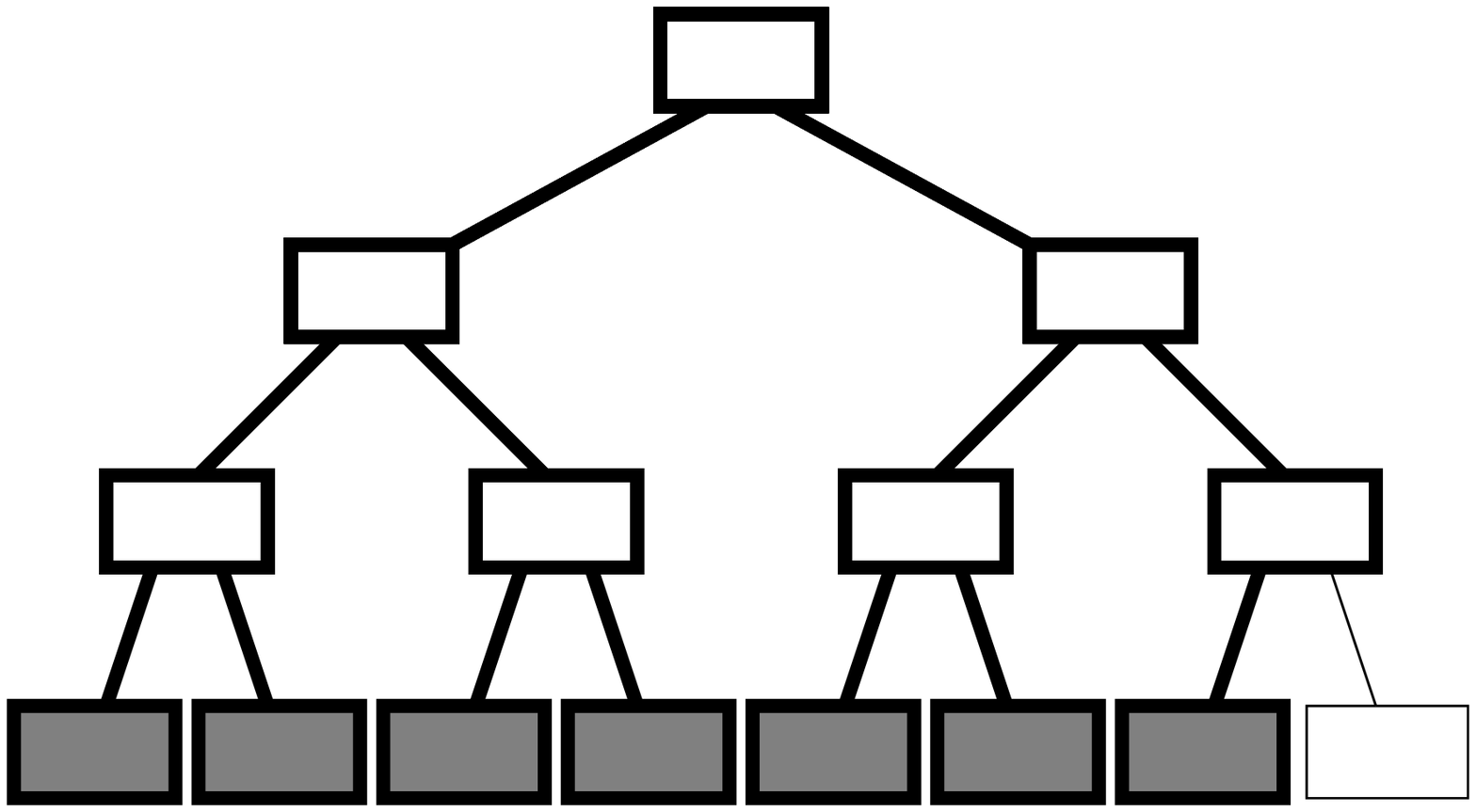} &
\includegraphics[width=0.23\textwidth]{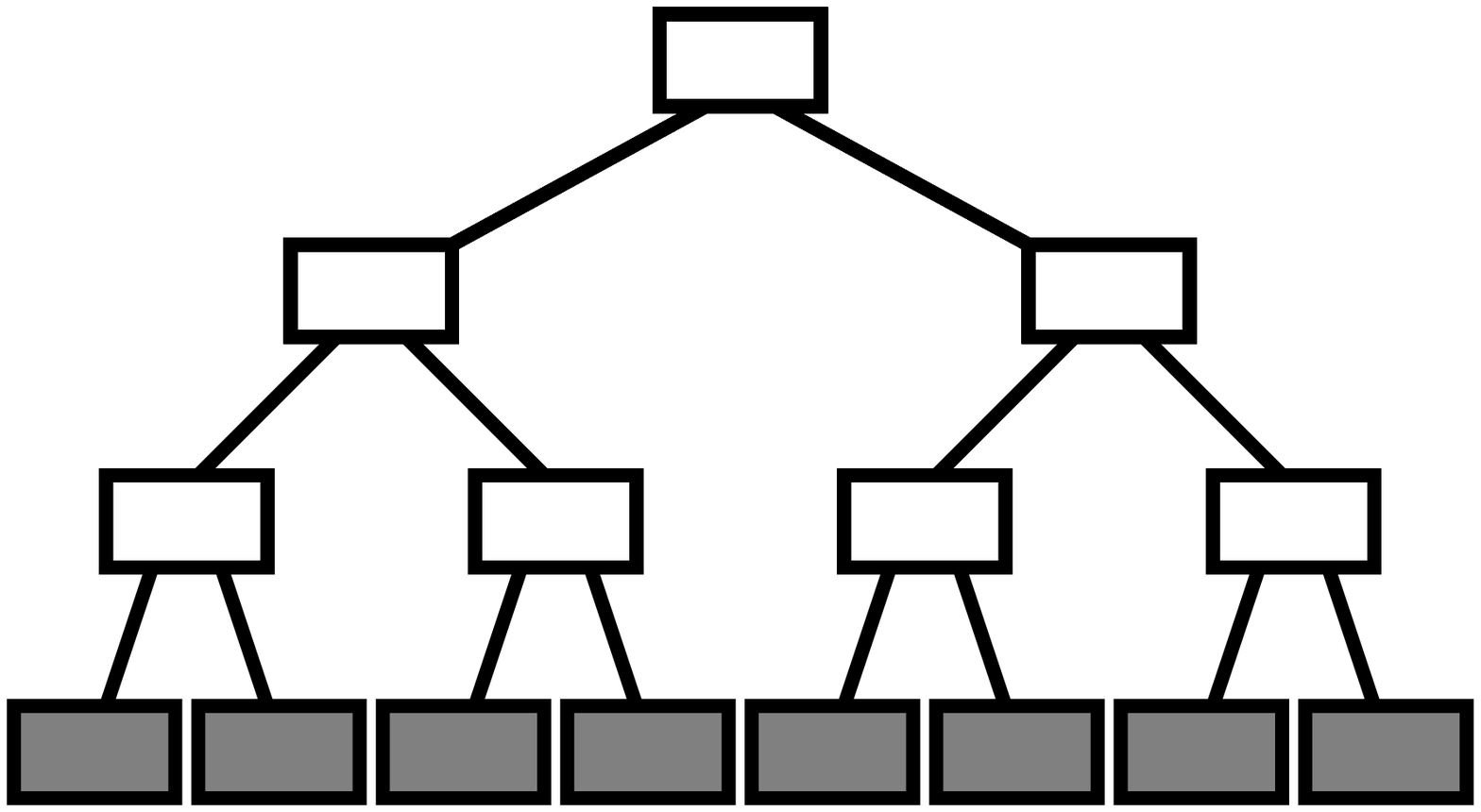}
\end{tabular*}$
\caption{The iterations of LDS on a search tree of depth $3$.}
\label{lds-iterations}
\end{figure}

LDS \cite{LDS} explores first those parts of the search 
tree that have a small number of discrepancies.
In each iteration of LDS the number of allowed discrepancies is incremented.
Fig.~\ref{lds-iterations} illustrates the iterations of LDS.
LDS has some redundancy since it only sets an upper bound on the allowed number of discrepancies.
Therefore, in iteration $x$, LDS examines the paths from previous iterations again.

\subsubsection{Improved Limited Discrepancy Search (ILDS)}
\begin{figure}[ht]
\centering
$\begin{tabular*}{\textwidth}{@{}@{\extracolsep{\fill}} c c c c @{}}
~\\
\multicolumn{1}{l}{\mbox{\scriptsize 1}}  &
\multicolumn{1}{l}{\mbox{\scriptsize 2}}  &
\multicolumn{1}{l}{\mbox{\scriptsize 3}}  &
\multicolumn{1}{l}{\mbox{\scriptsize 4}} \\ [-0.5cm]
\includegraphics[width=0.23\textwidth]{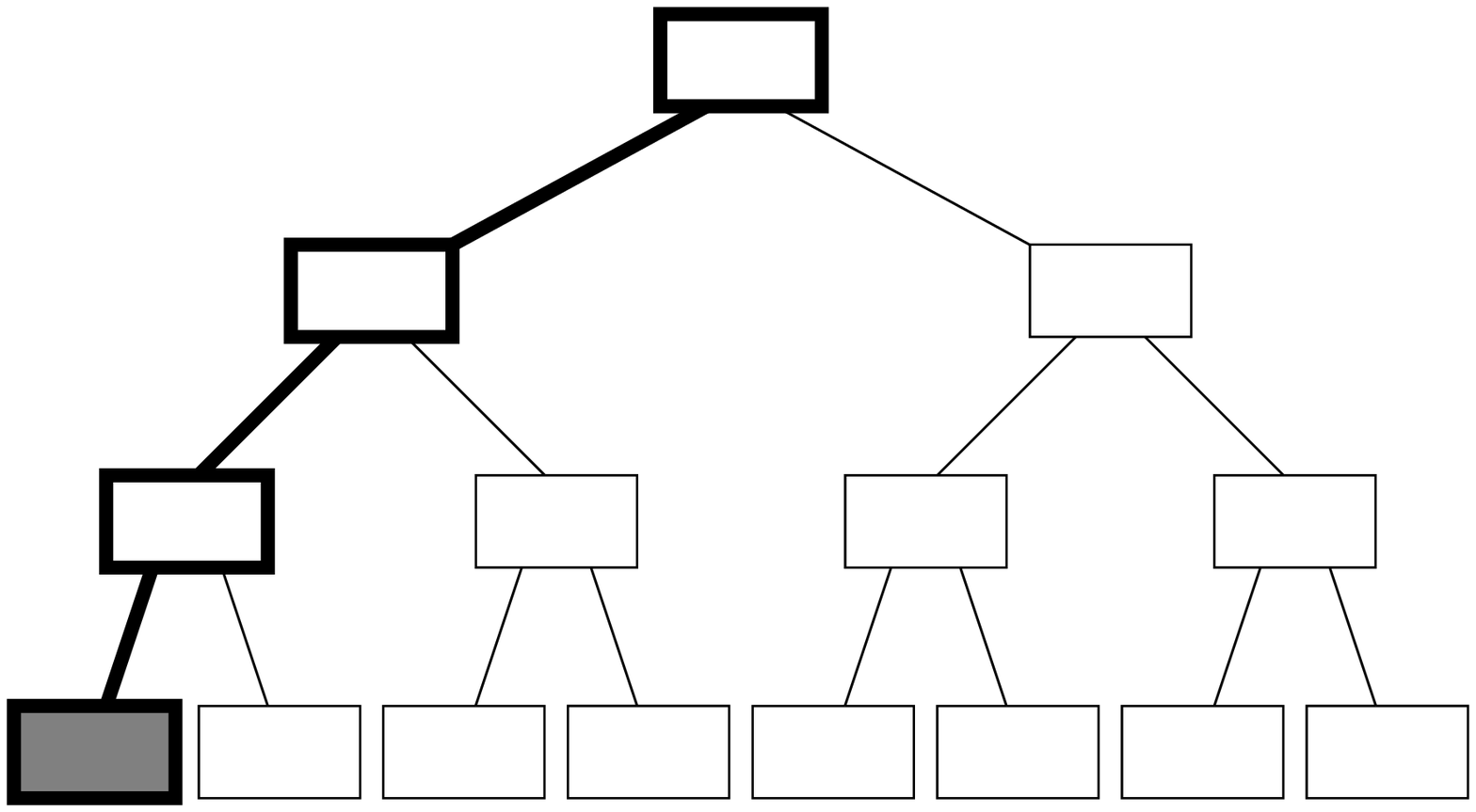} &
\includegraphics[width=0.23\textwidth]{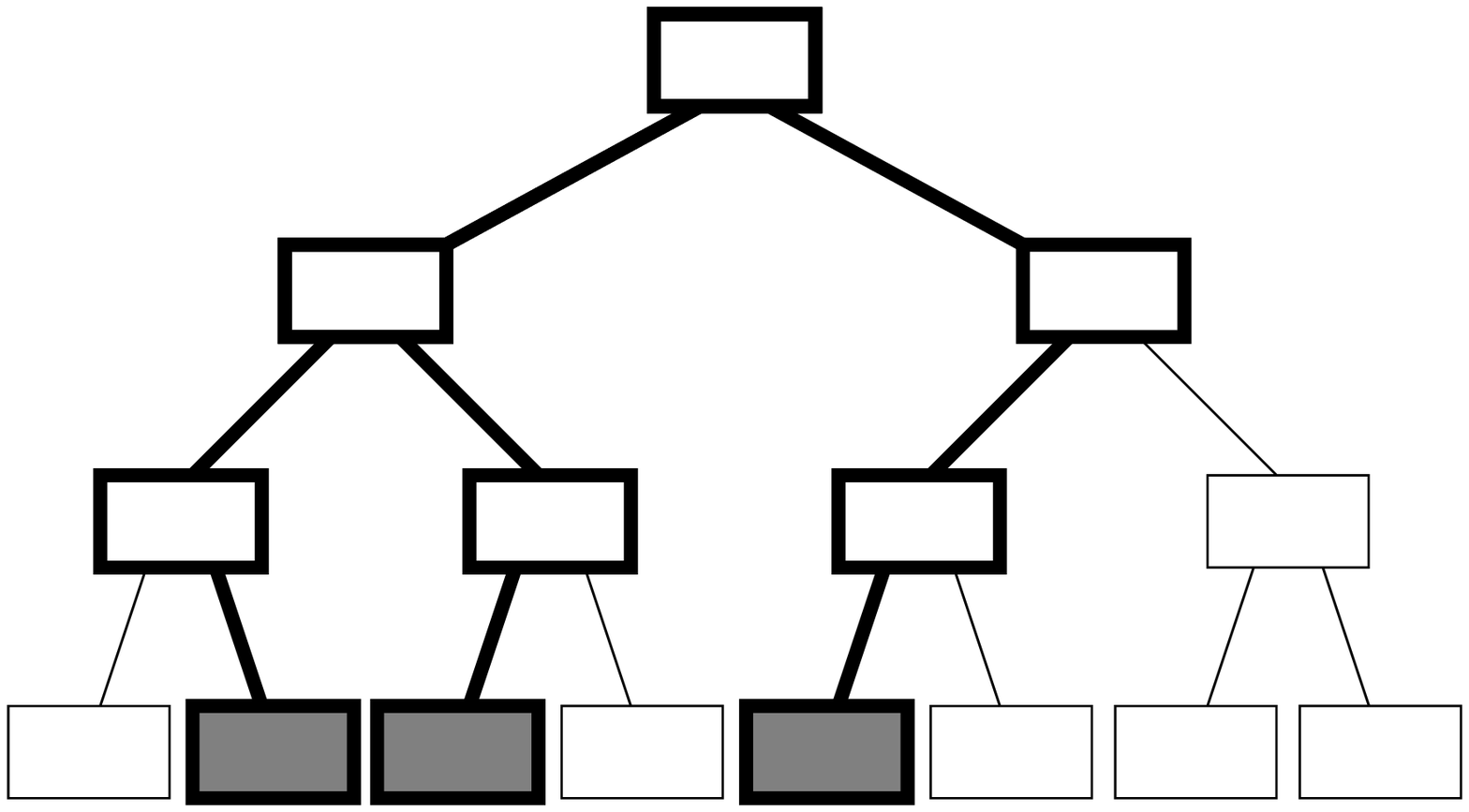} &
\includegraphics[width=0.23\textwidth]{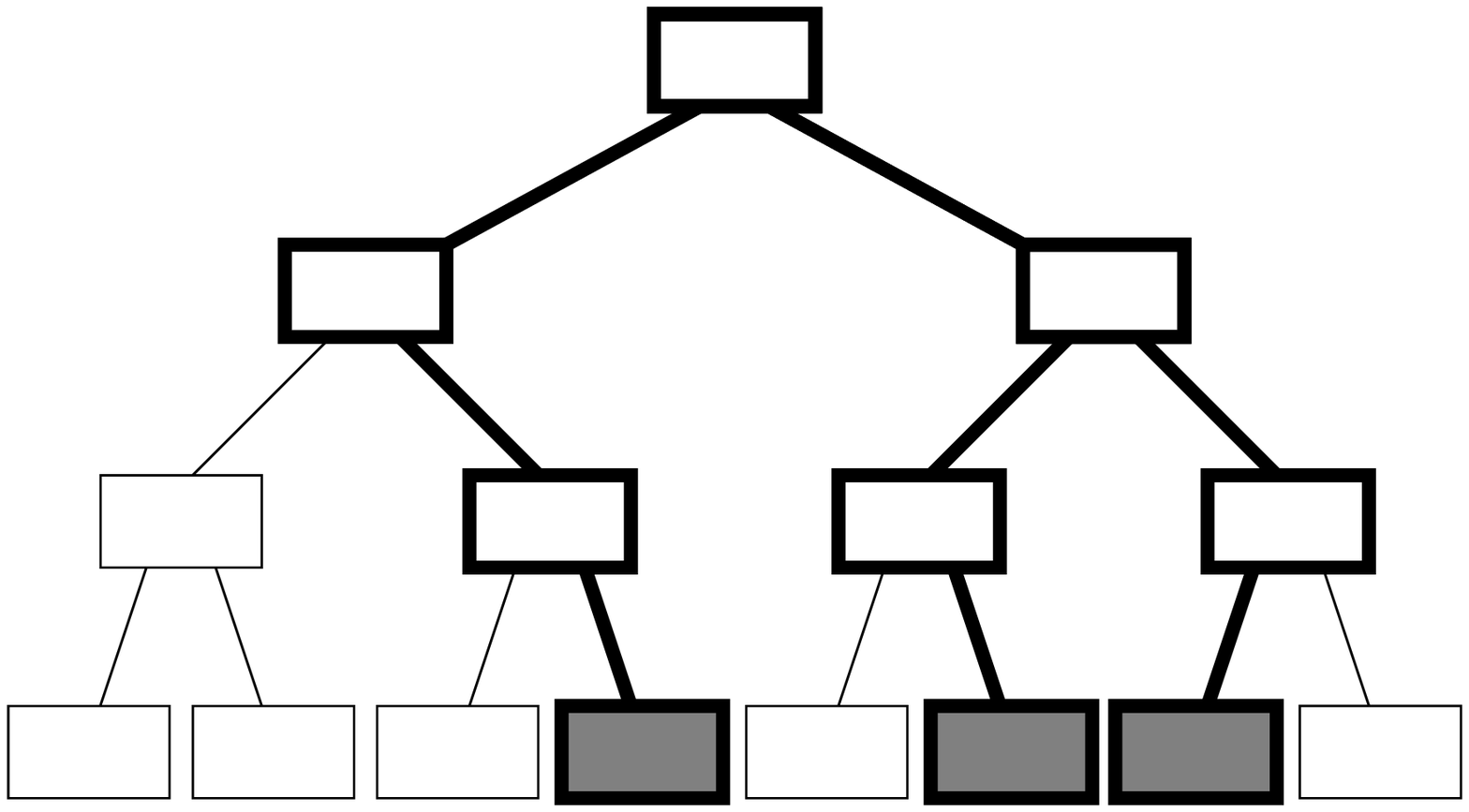} &
\includegraphics[width=0.23\textwidth]{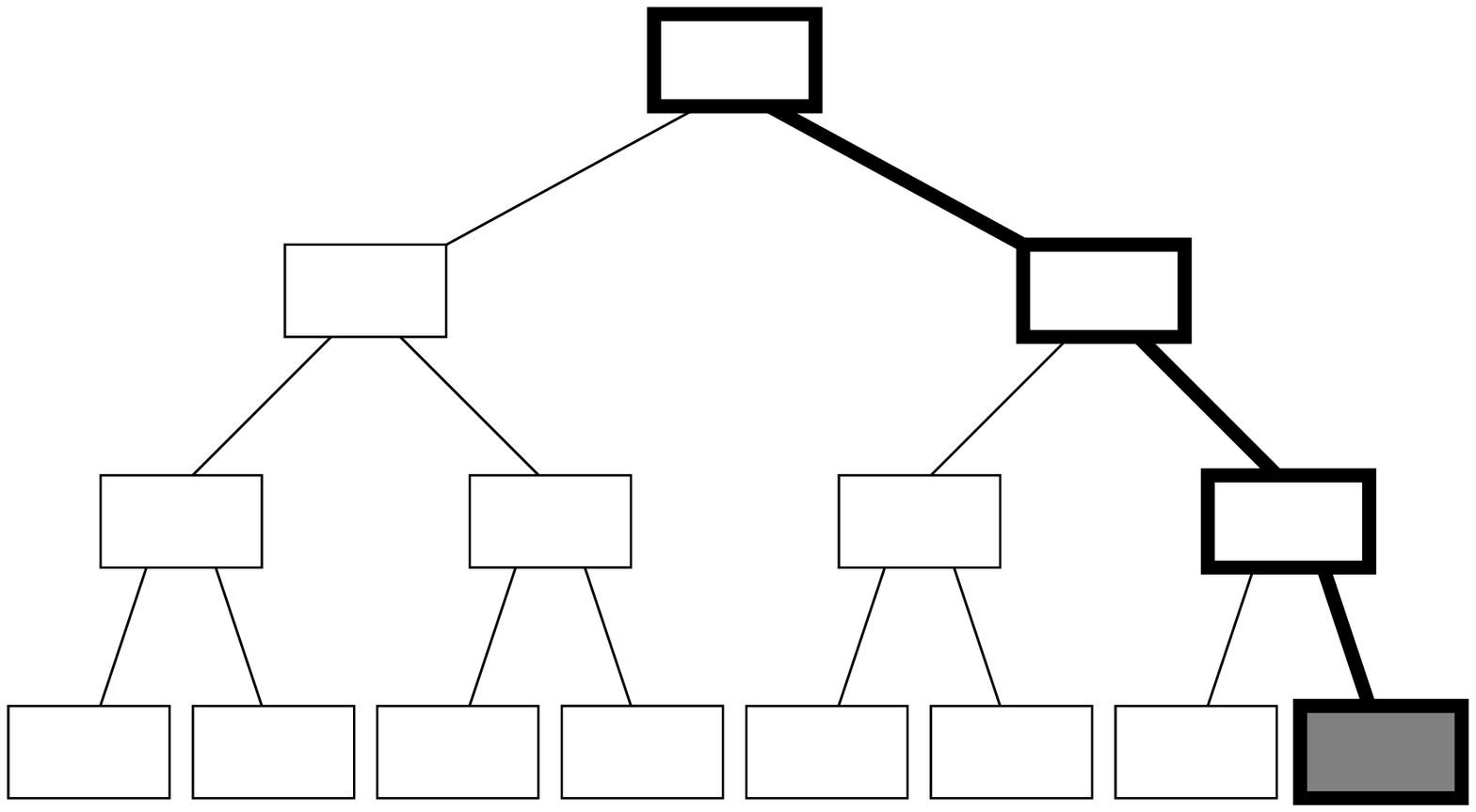} 
\end{tabular*}$
\caption{The iterations of ILDS on a search tree of depth $3$.}
\label{ilds-iterations}
\end{figure}

ILDS improves LDS by eliminating the redundancy.
This is achieved by providing a maximum search depth $d$ to the algorithm.
Given this depth, at any point during its execution, the algorithm keeps track of the remaining number of depths to be searched.
As a consequence, in each iteration $x$, only the paths with exactly $x$ discrepancies are explored (starting with $x=0$).
This way ILDS ensures that subtrees rooted at depth $d$ are explored only once.
All subtrees rooted at depth $d$ are searched using DFS.
The iterations of ILDS are shown in Fig. \ref{ilds-iterations}.

%
\subsubsection{Depth-bounded Discrepancy Search (DDS)}
\begin{figure}[ht]
\centering
$\begin{tabular*}{\textwidth}{@{}@{\extracolsep{\fill}} c c c c @{}}
~\\
\multicolumn{1}{l}{\mbox{\scriptsize 1}}  &
\multicolumn{1}{l}{\mbox{\scriptsize 2}}  &
\multicolumn{1}{l}{\mbox{\scriptsize 3}}  &
\multicolumn{1}{l}{\mbox{\scriptsize 4}} \\ [-0.5cm]
\includegraphics[width=0.23\textwidth]{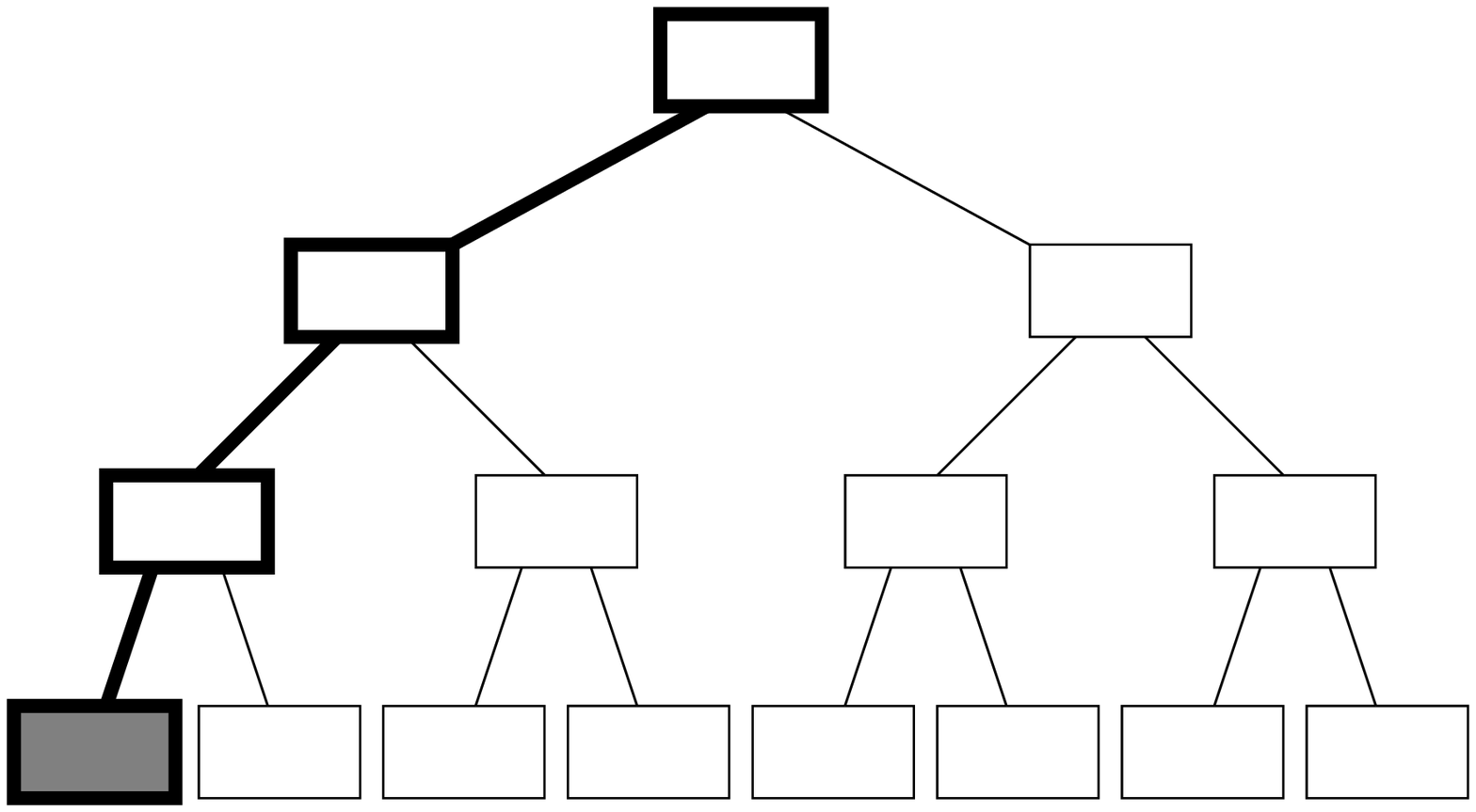} &
\includegraphics[width=0.23\textwidth]{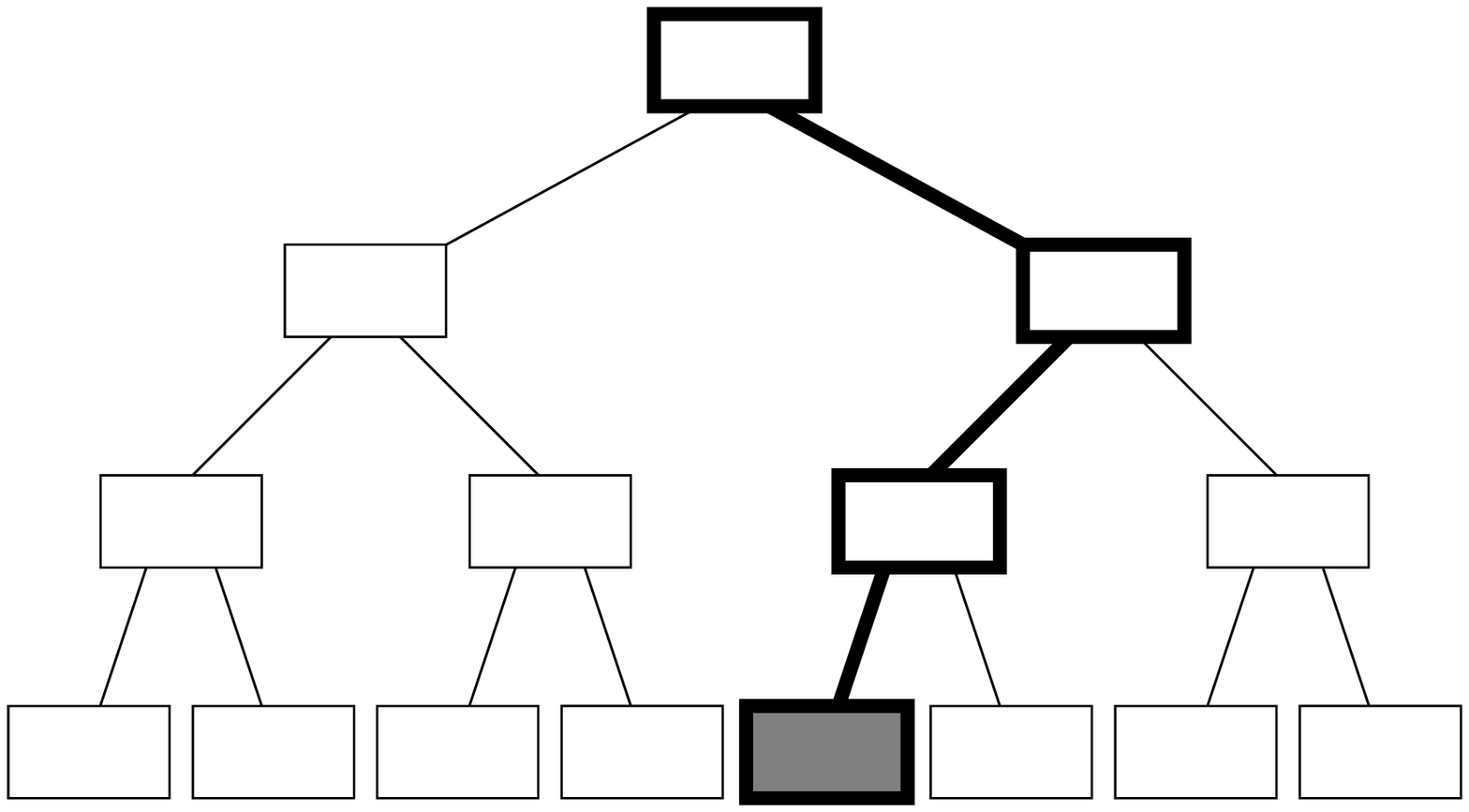} &
\includegraphics[width=0.23\textwidth]{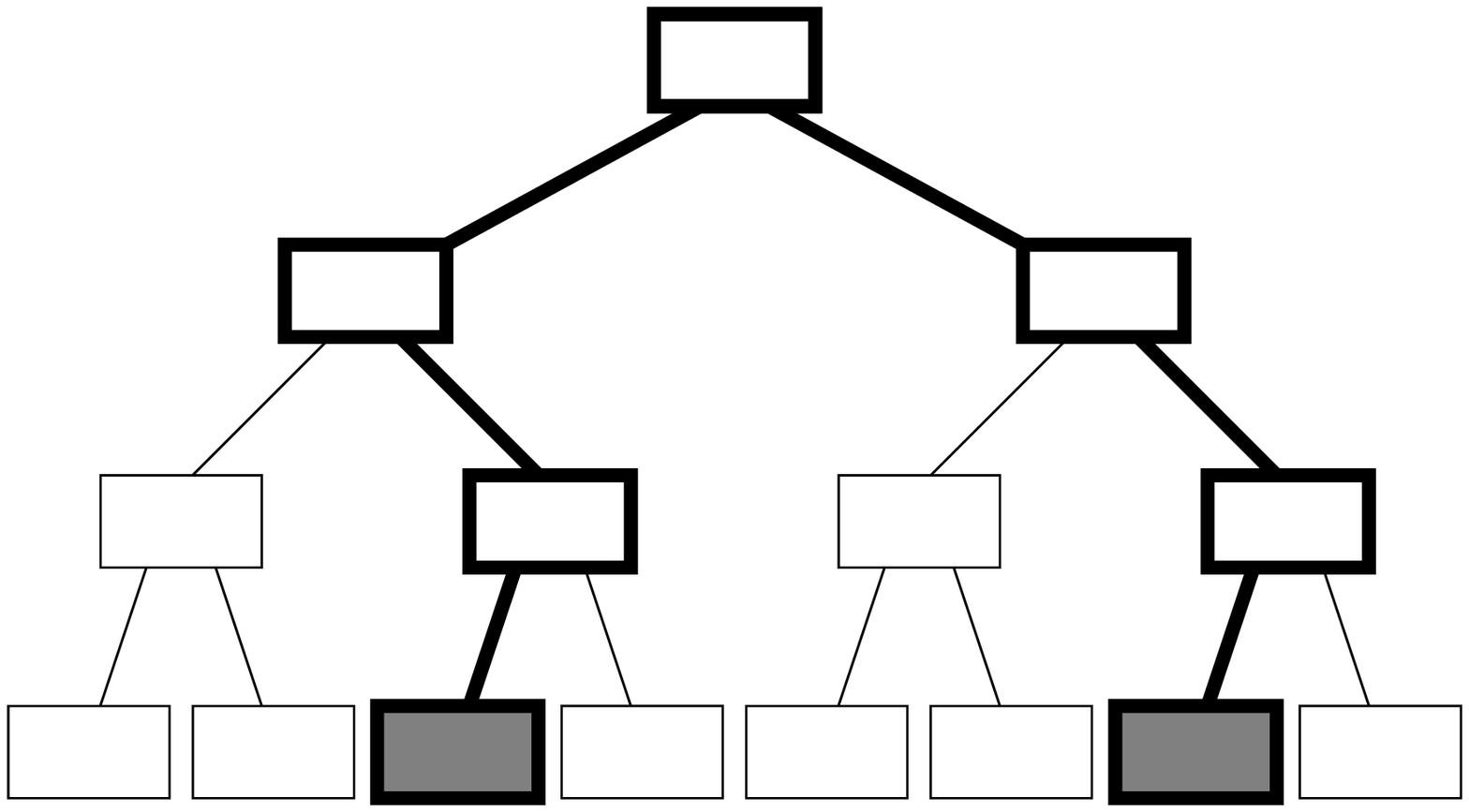} &
\includegraphics[width=0.23\textwidth]{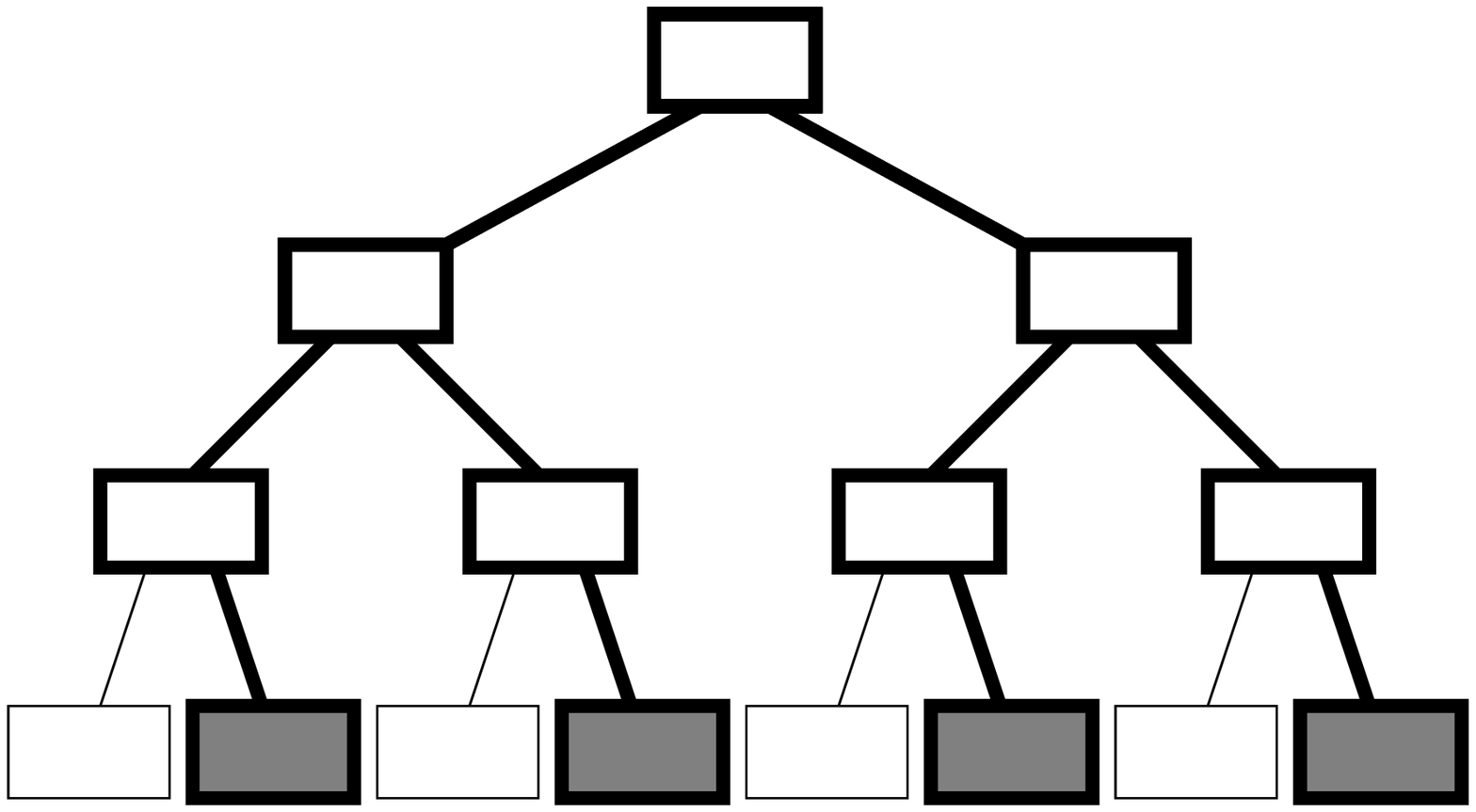}
\end{tabular*}$
\caption{The iterations of DDS on a search tree of depth 3.}
\label{dds-iterations}
\end{figure}

By incrementally increasing the maximum depth up to which discrepancies are allowed to 
occur, DDS~\cite{DDS} differs from (I)LDS. More specific, DDS visits in each iteration $i+1$, all branches at depth $d<i$, 
only the discrepancies at depth $d=i$, and no discrepancies are allowed for $d>i$.
Exploring the search tree this way also removes the requirement of specifying a maximum depth.
As can be seen in Fig. \ref{dds-iterations}, DDS explores paths with multiple right branches at the top of the search tree relatively early.
In specific cases where the direction heuristics are bad (heuristic probabilities are close to $0.5$ in case of a binary tree) near the root of the tree, but suddenly get very good (close to $1$) at a certain depth, it is useful to introduce multiple discrepancies at the top of the search tree early.

\subsection{Advanced Limited Discrepancy Search}
The papers describing LDS~\cite{LDS}, ILDS~\cite{ILDS} and DDS~\cite{DDS} are 
precise on which leaf nodes are explored in each of the iteration stages.
Yet, apart from the pseudocode, the order in which leaf nodes are visited within a single iteration stage is not explicitly specified.
We will assume that the strategies are applied as described in the pseudocode.
For ILDS, DDS this means that, in each iteration, leaf nodes are explored from left to right.

\begin{figure}[ht]
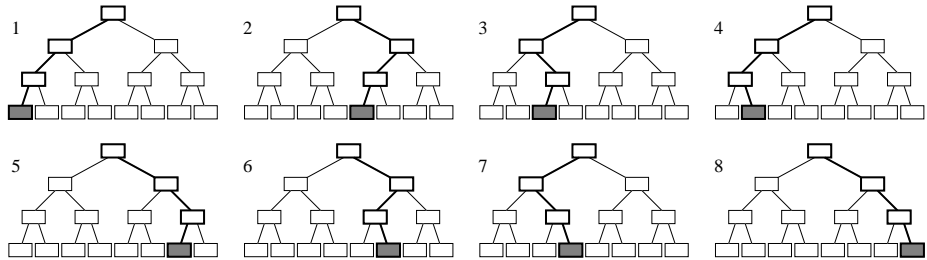

\centering
$\begin{tabular*}{\textwidth}{@{}@{\extracolsep{\fill}} c c c c @{}}
~\\
\multicolumn{1}{l}{\mbox{\scriptsize 1}}  &
\multicolumn{1}{l}{\mbox{\scriptsize 2}}  &
\multicolumn{1}{l}{\mbox{\scriptsize 3}}  &
\multicolumn{1}{l}{\mbox{\scriptsize 4}} \\ [-0.5cm]
\includegraphics[width=0.23\textwidth]{tree1.eps} &
\includegraphics[width=0.23\textwidth]{tree5.eps} &
\includegraphics[width=0.23\textwidth]{tree3.eps} &
\includegraphics[width=0.23\textwidth]{tree2.eps} \\ [0.3cm]
\multicolumn{1}{l}{\mbox{\scriptsize 5}}  &
\multicolumn{1}{l}{\mbox{\scriptsize 6}}  &
\multicolumn{1}{l}{\mbox{\scriptsize 7}}  &
\multicolumn{1}{l}{\mbox{\scriptsize 8}}  \\ [-0.5cm]
\includegraphics[width=0.23\textwidth]{tree7.eps} &
\includegraphics[width=0.23\textwidth]{tree6.eps} &
\includegraphics[width=0.23\textwidth]{tree4.eps} &
\includegraphics[width=0.23\textwidth]{tree8.eps} \\
\end{tabular*}$
\caption{The iterations of ALDS on a search tree of depth 3.}
\label{alds-iterations}
\end{figure}

By combining features from ILDS and DDS, a new search strategy can be created.
This search strategy uses the iterations of ILDS, while nodes within an iteration are visited according to DDS. 
More specific, nodes with the same number of discrepancies are visited from right to left.
We call this strategy \emph{Advanced Limited Discrepancy Search} (ALDS).
The iterations of this strategy are shown in Fig.~\ref{alds-iterations}.
Like ILDS, ALDS is only applied until a certain depth $d$, while subtrees rooted at $d$
are explored using DFS.

\vspace{-10pt}

\noindent This search strategy is inspired by an earlier experimental study on random 3-SAT 
instances, where we observed~\cite{side} 
two patterns regarding the goal node probabilities:
\begin{enumerate}
\item Leaf nodes reached with less discrepancies have a higher goal node probability.
\item For leaf nodes reached with the same number of discrepancies, those reached with 
discrepancies closer to the root have a higher goal node probability.
\end{enumerate}
\noindent
Notice that ALDS visits leaf nodes in the preferred order of these observations.

In order to compare the various search strategies, we propose a model to approximate the performance. 
In this model, the top of the search tree, until depth $d$, is visited using discrepancy search, while
all subtrees rooted at depth $d$ are visited by DFS. 
The depth $d$ at which DFS takes over from the discrepancy search is called the \emph{jump depth}.
So, using a jump depth $d$ would result in $2^d$ subtrees.
Discrepancy search ensures that promising parts of the search tree are explored first, while DFS 
searches the remaining subtree with minimal branching overhead.
Subtrees explored by DFS are considered leaf nodes in the discrepancy search.

If the size of subtrees rooted at depth $d$ is substantial, the cost of traversing a subtree is much higher than the overhead of 
jumping from one subtree to another. Assuming that subtrees at the same depth do not differ much in size, the cost of finding a goal
node can be approximated by the number of subtrees one expects to explore. For search trees with a single solution,
the expected cost can be computed using the $\pgoal(v)$ values of the nodes at the jump depth:

\def \order {\mathrm{order}}
\def \subtrees {\mathrm{subtrees}}
\def \egoal {E_{\mathrm{goal}}}

\vspace{-10pt}
\begin{equation}
\egoal (d) =\frac{1}{2^d} \cdot \sum_{v=1}^{2^d} \big(\order(v) \cdot \pgoal(v) \big)
\end{equation}

For a given depth $d$, subtrees are numbered chronologically from 1 to $2^{\!d}$ and $\order(v)$ denotes the 
index at which subtree $v$ will be visited in the specific search strategy.
For example, the summation for ALDS using the goal node probabilities in Fig.~\ref{prob-tree} is:

\vspace{-10pt}
\begin{equation*}
\frac{1}{8} \cdot
(1 \cdot .504 + 4 \cdot .056 + 3 \cdot .126 + 7 \cdot .014 +
2 \cdot .216 + 6 \cdot .024 + 5 \cdot .054 + 8 \cdot .006)
\end{equation*}
\vspace{-10pt}

The table below shows the $\egoal(3)$ values based on the Fig.~\ref{prob-tree} probabilities:
\vspace{-10pt}
\begin{table}[ht]
\centering
\begin{tabular}{ @{\hspace{15pt}}  l@{\hspace{15pt}} | @{\hspace{15pt}}c@{\hspace{15pt}}  @{\hspace{15pt}}c@{\hspace{15pt}}
				@{\hspace{15pt}}c@{\hspace{15pt}}	@{\hspace{15pt}}c@{\hspace{15pt}} }
 & DFS & ILDS & DDS & ALDS \\
\hline \hline
$\egoal(3)$ & $0.3375$ & $0.31225$ & $0.26375$ & $0.26225$
\end{tabular}
\end{table}
\vspace{-10pt}

\noindent
In case of multiple goal nodes, the 'expected' cost can be computed for a set of instances. Solve the instances
using a search strategy and determine the average number of subtrees visited at depth $d$. Divide the average number by $2^{\!d}$ to
obtain the average cost. We will denote this alternative by $\egoal^*(d)$.

\section{Experiments}
\label{sec:results}
Two types of results are presented in this section: theoretical and experimental results.
The theoretical results are based on a probabilistic model of heuristic tree search.
Experiments have been performed with the look-ahead SAT solver {\sf march}~\cite{side}, the fastest solver on random $3$-SAT benchmarks.

We compare several discrepancy-based search strategies on the theoretical model and on a dataset of random $3$-SAT formulae.
Additionally, two experiments were performed to determine how much ALDS could be improved.

\subsection{Theoretical results}
\label{section-theory}
Based on the increasing heuristic probability observation~\cite{side} we created a model with just one goal node.
In this model we assign the heuristic probability as follows (based on the observed $\pheur(v)$ values in ~\cite{side}):
\begin{equation}
\pheur(v) =  0.56 + 0.015 \cdot \depth(v)
\label{theory}
\end{equation}
Each leaf node $v$ at depth $12$ is assigned a goal node probability $\pgoal(v)$ using the equations described in Section \ref{section-heuristic}.
This means $\pgoal$ is calculated by multiplying the heuristic probabilities of the left and right children leading to that leaf node, starting at the root with $\pheur($root$) = 1$.
So, similar to the tree in Fig.~\ref{prob-tree}, only using a much larger tree.

In practice search trees  contain multiple goal nodes, but no generality is lost by putting just one goal node in the search tree of our model.
This will only cause expected cost to find a goal node and other numerical results to be a little larger, but this will not favour any search strategy in particular.
The difference in numerical results is acceptable because this model is only used to compare the search strategies to each other.

\begin{figure}[ht]
\centering
\input{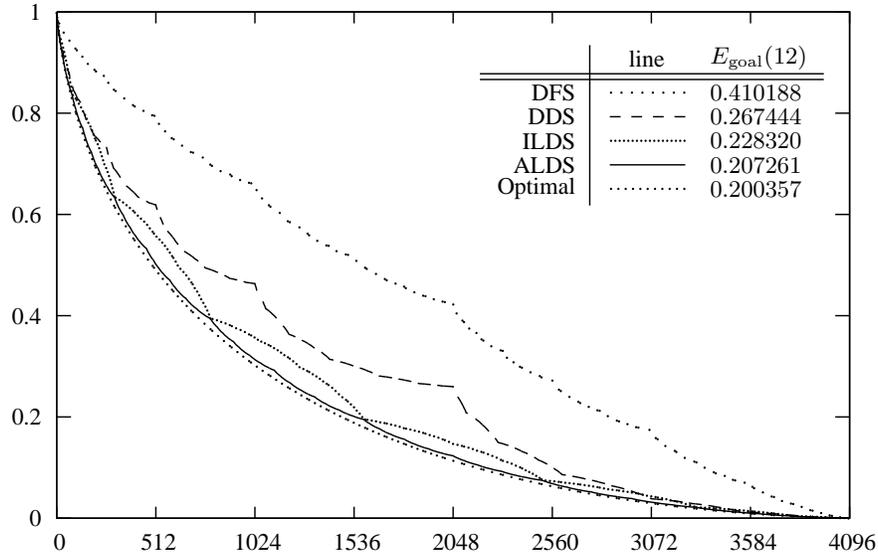}
\caption{Theoretical fraction of instances unsolved versus number of leaf nodes explored on a model search tree of depth $12$ containing a total of $4096$ leaf nodes.}
\label{theory-linear-graph}
\end{figure}

Because the goal node probability of each subtree is defined by our own model, the optimal order in which to search the subtrees is to go from high to low goal probability.
The expected fraction of the tree that has to be searched before a goal node is reached is the area below each graph (see Fig.~\ref{theory-linear-graph}).
ALDS performs best on the model.
In addition, the difference between ALDS and the optimal search order is quite small.

\subsection{Satisfiability results}
For the experimental results we used the look-ahead SAT solver {\sf march}~\cite{side} with the 
recursive weight heuristic $w_3^{\times}$ as direction heuristic (see Section~\ref{sec:rwh}).
The dataset for the experiments consisted of $20146$ satisfiable random 3-SAT instances with $350$ variables
and $1491$ clauses.
The clauses-to-variables ratio is $4.26$, which is the ratio where the probability of generating a satisfiable instance is about 50\%, 
known as the phase transition density.

For each instance, the complete search tree was explored to find out which of the $4096$ subtrees at depth $12$ contained solutions.
A depth of $12$ was chosen to keep the data compact enough for practical use, but still perform discrepancy search on a significant part of the search tree.
On average each instance contained $17.2$ satisfiable subtrees.

\begin{figure}[ht]
\centering
\input{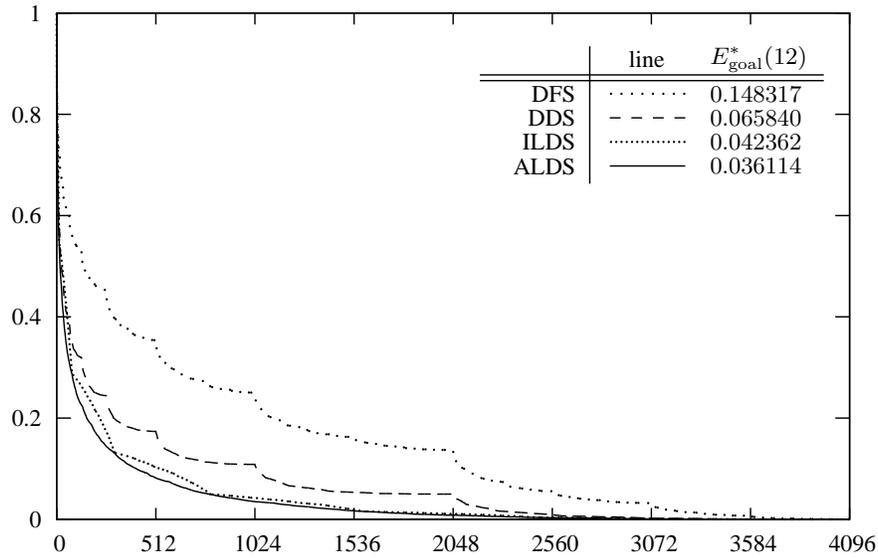}
\caption{Fraction of instances unsolved versus explored search space, expressed by the number of subtrees explored on depth $12$, for the 3-SAT instances.}
\label{pfirst-graph}
\end{figure}

Results from the experiments are shown in Fig.~\ref{pfirst-graph}.
The vertical axis shows the fraction of problems that has not been solved yet.
The horizontal axis displays the number of subtrees that have been explored.
Compared to the theoretical results, these lines decrease faster.
This can be explained by the fact that there is an average of $17.2$ goal nodes per instance in this experiment compared to a single goal node in the theoretical model. 
The $\egoal^*(12)$ values correspond to the area below the graphs.

\subsection{Analysis}

\begin{figure}[ht]
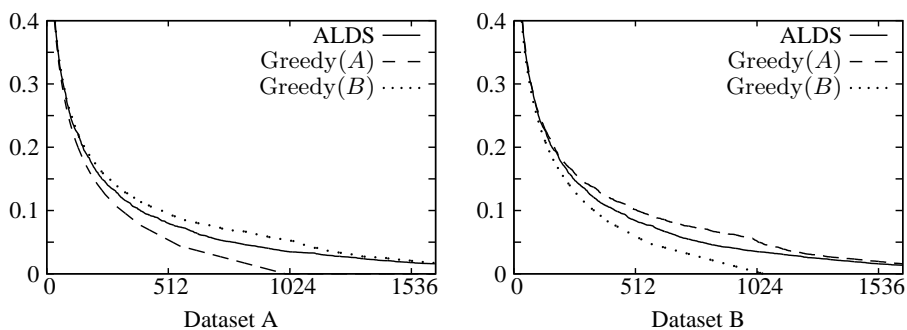

\centering

$\begin{tabular*}{\textwidth}{c c}
\input{Pfirst-greedy1.tex} &
\input{Pfirst-greedy2.tex} \\
Dataset A & Dataset B
\end{tabular*}$
\caption{Analysis of Greedy strategies versus ALDS.
Each of the pictures represents a distinct half of the dataset.
The Greedy order used to create the top line in one picture is the order used to create the lower line in the other picture.}
\label{greedy-graph}
\end{figure}

Similar to Section \ref{section-theory}, we want to demonstrate that ALDS performs close to optimal 
on the dataset of random 3-SAT instances. Yet, due to multiple satisfying subtrees per instance, 
it is hard to determine the performance of the optimal search strategy.
To approximate the optimal search strategy, we construct a \emph{Greedy} search strategy.
The Greedy search strategy is introduced in \cite{side}, and is constructed as follows:
\begin{itemize}
\item Select the subtree in which most instances from the dataset have at least one solution.
This subtree is next to be visited in this specific Greedy search strategy.
\item Remove from consideration all the instances in which the selected subtree has at least one solution.
\item Repeat above steps until all instances are removed from consideration.
\item The subtrees that have not been ordered yet, are placed at the end in ALDS order.
\end{itemize}
The construction of the Greedy search strategy requires a set of instances as input.
\def \greedy {\mathrm{Greedy}}
We let $\greedy(S)$ denote the Greedy search strategy that has been constructed with input set $S$.
Any Greedy search strategy will perform very well for the given input set.
To determine whether or not Greedy could actually make a good \emph{generalized} search strategy, 
the dataset has been split into two parts, which we will call part $A$ and part $B$.
When $\greedy(A)$ is applied to part $A$, the $\egoal^*(12)$ performance is very strong, as expected.
However, when $\greedy(A)$ is applied to part $B$ of the dataset, the result is worse than 
using the ALDS search method.
This can be observed in Fig.~\ref{greedy-graph}.

Furthermore, the $\greedy(S)$ lines on dataset $S$ in Fig.~\ref{greedy-graph} reach the horizontal axis 
after about $1024$ of the $4096$ subtrees. This is due to the limited size of the experiment.
On a hypothetical collection of all possible random $3$-SAT instances, this line could not reach 
the horizontal axis this quickly.
To summarize: the Greedy search strategy benefits greatly from the limited size of the dataset.
Yet, it seems unlikely that a greedy search strategy  can be converted into a generalized search strategy 
in such a way that it will outperform ALDS.

\vspace{5pt}

In our second analysis of the performance of ALDS, we experimented with various linear models for the 
$\pheur(v)$ values, see (\ref{model}). Our motivation to use this linear model is that we observed that the model
(\ref{theory}) is an accurate approximation of the $\pheur(v)$ values using {\sf march} with the $w^*_3$ heuristic
on random 3-SAT formulae.
%
\begin{equation}
\pheur(v) = y + x \cdot \mathrm{depth}(v)
\label{model}
\end{equation}
%
\noindent
We used these linear models to construct search strategies:

\vspace{-7pt}
\begin{itemize}
\item Compute for all 4096 subtrees at depth 12 the $\pgoal(v)$ values
\item Visit the subtrees based on the $\pgoal(v)$ values from high to low
\end{itemize}
\vspace{-7pt}

\noindent The objective is to determine which linear model results in the smallest $\egoal^*(12)$ value.  
Because we only use the  $\pgoal(v)$ values to determine the order of the subtrees, parameter $y$
can be fixed. We set it to $y=0.56$ as in (\ref{theory}). The table below shows the results:

\begin{table}[ht]
\vspace{-10pt}
\begin{tabular}{ c  @{\hspace{5pt}} | @{\hspace{5pt}} c @{\hspace{10pt}} c @{\hspace{10pt}} c 
				@{\hspace{10pt}} c @{\hspace{10pt}} c @{\hspace{10pt}} c @{\hspace{10pt}} c}
$x$ &  $0.0050$ &  $0.0055$ & $0.0060$ & $0.0065$ &  $0.0070$ &  $0.0075$ &  $0.0080$   \\
\hline \hline 
$\egoal^*(12)$     & $0.035331$    & $0.035325$     & $0.035327$ & $0.035340$ & $0.035341$ & $0.035372$ & $0.035390$
\end{tabular}

\begin{tabular}{ c  @{\hspace{5pt}} | @{\hspace{5pt}} c @{\hspace{10pt}} c @{\hspace{10pt}} c 
				@{\hspace{10pt}} c @{\hspace{10pt}} c @{\hspace{10pt}} c @{\hspace{10pt}} c}
$x$ &  $0.0090$ &  $0.0100$ & $0.0110$ & $0.0120$ &  $0.0130$ &  $0.0140$ &  $0.0150$   \\
\hline \hline 
$\egoal^*(12)$     & $0.035434$    & $0.035447$     & $0.035561$ & $0.035735$ & $0.035840$ & $0.035974$ & $0.036057$
\end{tabular}
\vspace{-5pt}

\end{table}

Two interesting things can be concluded from these experiments. First, the optimal value for the $x$ parameter ($0.0055$) is much
smaller than the value that matches the observed data ($0.0150$). Second, the optimal $\egoal^*(12)$ value ($0.03525$) is 
hardly smaller than the cost of ALDS ($0.036114$). It will be difficult to construct a search strategy based on 
a linear model that will outperform ALDS, because the overhead required to implement a more complex strategy will probably
be more costly than the small reduction in $\egoal^*(12)$ can compensate.

\section{Conclusion}
\label{sec:conclusions}

We introduced the recursive weight heuristic, a branching heuristic for look-ahead SAT solvers.
The solver {\sf march} equipped with this heuristic performs stronger than any other solver
on unsatisfiable random 3-SAT formulae. It won the gold medal in this category at the SAT
2009 competition.

To capitalize on the recursive weight heuristic on satisfiable instances as well, we presented
advanced limited discrepancy search (ALDS). Theoretical and practical results show that
ADSL outperforms alternative complete search strategies on satisfiable random 3-SAT 
formulae. We show that on these instances, ALDS in combination with the recursive weight
heuristic traverse the search tree in a near-optimal order.

\bibliography{ALDS}
\bibliographystyle{splncs}

\end{document}